\documentclass[useAMS,usenatbib]{mnras}
\usepackage{graphicx}
\usepackage{color}
\usepackage{amssymb}
\usepackage{epsfig}

\usepackage{comment}

\title[On the Radio Image of Relativistic Jets --- I]
{On the Radio Image of Relativistic Jets --- I: Internal Structure, Doppler Boosting, and Polarisation Maps}
\author[A.~V.~Chernoglazov, V.~S.~Beskin and V.~I.~Pariev]
{A.~V.~Chernoglazov$^{1,2,3}$\thanks{E-mail: alexander.chernoglazov@gmail.com (ACh)}, 
V.~S.~Beskin$^{2,3}$ and
V.~I.~Pariev$^{2}$
\\
$^{1}$University of New Hampshire, 105 Main St., Durham, NH 03824, USA\\
$^{2}$P.N.Lebedev Physical Institute, Leninsky prospekt 53, Moscow 119991, Russia \\
$^{3}$Moscow Institute of Physics and Technology, Institutsky per.~9, Dolgoprudny 141700, Russia \\
}

\begin{document}

\date{Accepted. Received; in original form}

\pagerange{\pageref{firstpage}--\pageref{lastpage}} \pubyear{2018}

\maketitle

\label{firstpage}

\begin{abstract}
In this first paper from forthcoming series of works devoted to radio image 
of relativistic jets from active galactic nuclei the role of internal structure 
of a flow is discussed. We determine the radial profiles of all physical values 
for reasonable Michel magnetisation parameter 
$\sigma_{\rm M}$ and ambient pressure $P_{\rm ext}$. Maps of  Doppler boosting 
factor $\delta$ and observed directions of linear polarisation of synchrotron emission 
are also constructed. 
\end{abstract}



\section{Introduction}
Recent progress in high angular resolution VLBI observations of relativistic
jets outflowing from active galactic nuclei~\citep{Mertens, MOJAVE_XIV} allows 
us to investigate directly their internal structure. In particular, 
the observations give 
us direct information about the dependence of the jet width $r_{\rm jet}(l)$ on the 
distance $l$ from the ''central engine''. 

Progress in VLBI observations allows us also to relate to more detailed information 
from the theory of relativistic jets. In spite of the wide variety of analytical and
numerical models on jets acceleration and confinement  ~\citep{ Vlahakis03, 
McKinney06, Komissarov07, Tchekhovskoy_11, McKinney12, PC15} considering different 
solutions for jets shapes, there is no common point of view on the internal structure 
of relativistic jets. 

In a number of works \citep{Parievetal, LPG, Porth_etal11, Fendt_conf}
observable maps of synchrotron polarisation and Faraday rotation for 
different models of relativistic jets in AGNs are suggested. The distribution 
of these quantities contains the information about magnetic field geometry,
which is mostly toroidal. On the other hand, no predictions about Doppler 
boosting factor distribution have been done. The Doppler boosting factor 
was present in formulas, but its importance was not analysed. The ignorance 
of the Doppler boosting factor map was based on smallness of variation of
toroidal velocity component ($ v_{\varphi}\ll c $), and it used to be impossible
to measure the velocity directly. Nevertheless, it has been recently 
discovered~\citep{Mertens} via wavelet analysis that the radiating plasma 
inside the jet in the nearest radio galaxy M87 rotates slowly.

Another vital difference of our consideration from previous ones is the 
ability to consider all parts of the jet self-consistently. Below we use 
our model of quasi-cylindrical flow with zero total electric current
inside the jet submerged into warm external gas in rest with reasonable 
thermal pressure: we do not assume neither external uniform magnetic field 
nor electromagnetic discontinuities at the jet boundary which were 
appearing in many works (see, e.g.,~\citealt{Lyu09, Marti}). This 
model gives us possibility to construct the most comprehensive radio map 
of FRII-type jet.

The problem of the pressure-balanced jet without current sheet
was considered by \citet{GFCL} (see also \citet{Kim1, Kim2}). But their 
consideration was based on modification of \citet{Lundquist} force-free solution,
where the gas pressure was not related to density by a reasonable equation of state
(the same approximation was used by \citet{Marti}). In addition, it was assumed that
the toroidal component of magnetic field is of the same order as the poloidal one 
 $B_z \sim B_{\varphi}$ instead of $B_z \ll B_{\varphi}$ in our model.

This paper is the first one in the forthcoming series of works where we will
investigate the footprints of internal structure of relativistic jets in 
their radio images. For this we determine the transverse profiles of main 
physical parameters of relativistic jets such as the number density of 
particles, $n_{\rm e}$, as well as toroidal and poloidal components of 
velocity ${\bf v}$ and magnetic field ${\bf B}$. Below we use semi-analytical 
1D cylindrical approach introduced by~\citet{Beskin97, BM00}. Further 
researches~\citep{LHAN99, BN06, BN09, Lyu09, NBKZ15} demonstrated that 
this simple approach allows us to describe principal properties of the 
internal structure of relativistic jets. For example, analytical asymptotic 
solutions for hydrodynamical Lorentz factor $\gamma$

\begin{eqnarray}
\gamma & \approx & \frac{r_{\perp}}{R_{\rm L}}, \qquad \qquad \, \, \,
R_{\rm c} > \frac{r_{\perp}^3}{R_{\rm L}^2},
\label{gammaff} \\
\gamma & \approx & \left(\frac{R_{\rm c}}{r_{\perp}}\right)^{1/2}, \qquad 
R_{\rm c} < \frac{r_{\perp}^3}{R_{\rm L}^2}.
\end{eqnarray}

which was later reproduced numerically in 3D simulations~\citep{McKinney06,
Komissarov07, Tchekhovskoy_11}. Here $r_{\perp}$ is the distance from 
the rotation axis, $R_{\rm L} = \Omega/c$ is the radius of the light 
cylinder, and $R_{\rm c}$ is the curvature radius of magnetic surface.

In the first paper of the series of works we present maps (two dimensional
distributions) of the Doppler boosting factor 
\begin{equation}
\delta = \frac{\sqrt{1-v^2/c^2}}{1-v\cos\chi /c}
\label{delta}
\end{equation}
and unit vector of the wave electric field in linearly polarised synchrotron 
radiation on the circular cross sections of the jet. In expression~(\ref{delta}) 
$\chi$ is the angle between 
the velocity of a parcel of plasma and the line of sight. 
These maps are determined by the magnetic field and the velocity of the MHD flow 
in the jet only, and are not 
influenced by the distribution of the number density of relativistic emitting 
particles, which is a major unknown ingredient in the jet models. More detailed 
simulations of observable radio images of jets on the sky, including the impact
of the modelling of radiating particles distributions and spectra, will be 
presented in the forthcoming paper.

\section{1D cylindrical Grad-Shafranov equation}
\label{GS-section}

Basic equations describing the structure of relativistic axisymmetric stationary  
flows within Grad-Shafranov approach were formulated about forty years ago (e.g., see
\citealt{Ardavan}). This approach allows us to determine the internal structure 
of axisymmetric stationary jets knowing in general case five ``intergals of motion'', 
which are energy $E(\Psi)$ and angular momentum $L(\Psi)$ flux, electric potential 
related to angular velocity $\Omega_{\rm F}(\Psi)$, entropy $s(\Psi)$, and the
particle-to-magnetic flux ratio $\eta(\Psi)$. All these quantities have to be constant 
along magnetic surfaces $\Psi = \mathrm{const}$. In particular, it was shown that a 
jet with total zero electric current can exist only in the presence of the external 
media with finite pressure $P_{\rm ext}$. Thus, it is the ambient pressure $P_{\rm ext}$
that determines the transverse dimension of astrophysical jets.

As was shown by~\citet{Beskin97, BM00}, for cylindrical flow it is convenient to 
reduce one second-order Grad-Shafranov equation to two first-order ordinary 
differential equations for magnetic flux $\Psi(r_{\perp})$ and poloidal Alfv\'enic 
Mach number ${\cal M}(r_{\perp})$
\begin{equation}
{\cal M}^2 = \frac{4\pi\mu\eta^2}{n}.
\label{M2}
\end{equation}
Here $n$ is the number density in the comoving reference frame and $\mu$ is 
relativistic enthalpy. The first equation is the relativistic Bernoulli 
equation $u_{\rm p}^2 = \gamma^2 - u_{\varphi}^2 - 1$, where $u_{\rm p}$ and
$u_{\varphi}$ are the poloidal and toroidal components of four-velocity 
${\bf u}$ respectively. Replacements for $\gamma$,  $u_{\rm p}$, and 
$u_{\varphi}$ in the Bernoulli equation lead to the form
\begin{eqnarray}
\frac{{\cal M}^4}{64\pi^4 r_{\perp}^2}
\left(\frac{{\rm d}\Psi}{{\rm d}r_{\perp}}\right)^2 =
\frac{K}{r_{\perp}^2 A^2} - \mu^2\eta^2 \,.
\label{ap4}
\end{eqnarray}
Here
\begin{equation}
A = 1-\Omega_{\rm F}^2 r_{\perp}^2/c^2-{\cal M}^2
\end{equation}
is the Alfv\'enic factor, 
\begin{equation}
K= r_{\perp}^2(e')^2(A-{\cal M}^2)+{\cal M}^4 r_{\perp}^{2}E^2-{\cal M}^{4}L^2c^2 \,,
\label{ap5}
\end{equation}
and $e' = E-\Omega_{\rm F}L$. 
The second equation determines the Mach number ${\cal M}$
\begin{eqnarray}
\left[\frac{(e')^2}{\mu^2\eta^2}-1+\frac{\Omega_{\rm F}^2 r_{\perp}^2}{c^2}
-A\frac{c_{\rm s}^2}{c^2}\right]
\frac{{\rm d}{\cal M}^2}{{\rm d} r_{\perp} } =
\frac{{\cal M}^6L^2c^2}{A r ^3 \mu^2\eta^2}
\nonumber \\
+\frac{\Omega_{\rm F}^2 r_{\perp} {\cal M}^2}{c^{2}}
\left[2 - \frac{(e')^2}{A\mu^2\eta^2}\right]
+{\cal M}^2 \frac{e'}{\mu^2\eta^2}\frac{{\rm d}\Psi}{{\rm d}r_{\perp}}\frac{{\rm d}e'}
{{\rm d}\Psi} \label{rel-1} \\
+\frac{{\cal M}^2 r_{\perp}^2}{2c^2}\frac{{\rm d}\Psi}{{\rm d}r_{\perp}}
\frac{{\rm d}\Omega_{\rm F}^2}{{\rm d}\Psi}
-{\cal M}^2 \left(1-\frac{\Omega_{\rm F}^2 r_{\perp}^2}{c^2} + 2A\frac{c_{\rm s}^2}{c^2}
\right)
\frac{{\rm d}\Psi}{{\rm d} r_{\perp} }\frac{1}{\eta}\frac{{\rm d}\eta}{{\rm d}\Psi}
\nonumber \\
-\left[\frac{A}{n}\left(\frac{\partial P}{\partial s}\right)_n
+\left(1-\frac{\Omega_{\rm F}^2 r ^2}{c^2}\right)T\right]
\frac{{\cal M}^2}{\mu}\frac{{\rm d}\Psi}{{\rm d}r_{\perp}} \frac{{\rm
d}s}{{\rm d}\Psi} \,.\nonumber  
\end{eqnarray}
Here $T$ is the temperature, $c_{\rm s}$ is the sound velocity defined as
$c_{\rm s}^{2}=(\partial P/\partial n)|_{s}/m_{\rm p}$, 
$\mu=m_{\rm p}c^2 +c_{\rm s}^2/(\Gamma-1)$ is again the relativistic enthalpy,
$m_{\rm p}$ is the particle mass, and $P$ is the gas pressure. As a result, 
Bernoulli equation (\ref{ap4}) and equation (\ref{rel-1}) form the system of 
two ordinary differential equations for the Mach number ${\cal M}(r_{\perp})$ 
and the magnetic flux $\Psi(r_{\perp})$ describing cylindrical relativistic jets. 

As was already stressed, the solution depends on our choice of five integrals 
of motion. On the other hand, it is important to note that by determining the 
functions ${\cal M}^2(r_{\perp})$ and $\Psi(r_{\perp})$ one can find the jet 
radius $d_{\rm jet}$ as well as the profile of the current $I(r_{\perp})$, the 
particle energy, and the toroidal component of the four-velocity from the 
solution of a problem under consideration. In particular, as
\begin{eqnarray}
\frac{I}{2\pi} & = & \frac{L-\Omega_{\rm F}r_{\perp}^{2}E/c^2}
{1-\Omega_{\rm F}^{2}r_{\perp}^{2}/c^2-{\cal M}^{2}},
\label{p33}
\end{eqnarray}
the condition of the closing of the electric current within 
the jet $I(\Psi_{\rm tot})= 0$ can be rewritten as $L(\Psi_{\rm tot}) = 0$ and 
\mbox{$\Omega_{\rm F}(\Psi_{\rm tot}) = 0$} simultaneously, where $\Psi_{\rm tot}$
is the given total magnetic flux in the jet. 

For this reason we use the following expressions for these integrals
\begin{eqnarray}
L(\Psi) & = & \frac{\Omega_{0} \Psi}{4 \pi^2} \sqrt{1 - \frac{\Psi}{\Psi_{\rm 
tot}}},
\label{L0}  \\
\Omega_{\rm F}(\Psi)  & = & \Omega_{0} \, \sqrt{1 - \frac{\Psi}{\Psi_{\rm 
tot}}}.
\label{Om0}
\end{eqnarray}
In the vicinity of a rotation axis these integrals correspond to 
well-known analytical force-free solution for homogeneous poloidal
magnetic field (see~\citealt{MHD} for more detail). On the other 
hand, they both vanish at the jet boundary which guarantees the 
fulfilment of the condition $I(\Psi_{\rm tot}) = 0$. 

Here we want to stress the novel point in our present work. We recall that careful 
matching of a solution inside the jet with the external media was produced only
recently. The difficulty in doing the matching is due to very low energy density 
of the external media in comparison with the energy density inside the relativistic 
jet. For this reason, in most cases the infinitely thin current sheet was introduced 
at the jet boundary. Moreover, the external pressure was very often modelled by 
homogeneous magnetic field $ B_{\rm ext}^2/8 \pi = P_{\rm ext}$.

In contrast,~\citet{BCKN-17} presented an approach which is free of the 
difficulties mentioned above. Following this paper we consider relativistic jet 
submerged into unmagnetised external media with finite gas pressure $P_{\rm ext}$. 
Neither external magnetic field nor infinitely thin current sheet are assumed. 
We succeed in doing so due to the boundary conditions (\ref{L0})--(\ref{Om0}) at the jet 
boundary $r_{\perp} = d_{\rm jet}$. 

In addition, we assume that both magnetic field and flow velocity vanish at
the jet boundary $r_{\perp} = d_{\rm jet}$. As one can see from (\ref{p33}), 
our choice of $L(\Psi)$ and $\Omega_{\rm F}(\Psi)$ guarantees that toroidal
component $B_{\varphi} \propto I/r$ vanishes at the jet boundary.
One can find that the toroidal velocity $u_{\varphi}$ also vanished at the 
jet boundary due to algebraic relation
\begin{equation}
u_{\varphi}  =  \frac{1}{\mu\eta c r} \, \frac{(E-\Omega_{\rm F}L)
 \Omega_{\rm F}r^{2}/c^2-L{\cal M}^{2}}{1-\Omega_{\rm F}^{2}r^{2}/c^2-{\cal M}^{2}}.
\label{p35}
\end{equation}
On the other hand, using relation $n {\bf u}_{\rm p} = \eta {\bf B}_{\rm p}$ one 
can conclude that the conditions ${\bf u}_{\rm p} = 0$ and ${\bf B}_{\rm p} = 0$ 
can be compatible with one another for finite $n$ and $\eta$. For simplicity we 
consider here the case 
\begin{equation}
\eta(\Psi) = \eta = {\rm const}. 
\label{eta0}
\end{equation} 
In addition, we suppose that the flow remains supersonic up to the very 
boundary: ${\cal M}(d_{\rm jet}) > 1$. This supposition allows us to simplify 
our consideration. Indeed, in this case equation (\ref{rel-1}) has no
additional singularity at the Alfv\'enic surface $A = 0$ in the vicinity of the
jet boundary. Finally, in what follows we put
\begin{equation}
s(\Psi) = s = {\rm const}. 
\label{s0}
\end{equation}  
This nonzero value allows us to match magnetically
dominated flow to the external media with finite pressure. 
As was shown by~\citet{BCKN-17}, thermal effects 
play important role only in the very vicinity
of the jet edge, while the main properties of the jet depend
on the ambient pressure $P_{\rm ext}$ (the value of $s$ determines 
the width of the boundary layer).

Finally, energy integral $E(\Psi)$ is expressed as \citep{BO}
\begin{equation}
E(\Psi) = \Omega_{\rm F}L + \mu_{0} \eta \gamma(\Psi).
\end{equation}
Here $\gamma(\Psi)$ is the Lorentz factor at the base of the jet, 
and $\gamma(0) = \gamma_{\rm in}$ is the Lorenz factor on the jet 
axis. We require that $\gamma(\Psi_{\rm tot}) = 1$ meaning that 
the velocity vanishes at the jet boundary. Below for simplicity 
we chose the radial profile of the injection Lorentz factor as 
\begin{equation}
\gamma(\Psi)=\gamma_{\rm in} - (\gamma_{\rm in} - 1) \frac{\Psi}{\Psi_{\rm tot}}.   
\label{gammain}
\end{equation}

In addition to five integrals of motion, the system (\ref{ap4}), (\ref{rel-1}) 
requires two boundary conditions. The first one is the clear condition at the 
symmetry axis
\begin{equation}
\Psi(0) = 0.
\label{bound-cond}
\end{equation}
As to the second one, it can be found from the pressure balance at the
jet boundary
\begin{equation}
P(d_{\rm jet}) = P_{\rm ext}.
\label{bound}
\end{equation}
This procedure fully determines the solution of the problem (\ref{ap4}),
(\ref{rel-1}), and (\ref{bound-cond})--(\ref{bound}), and ensures its 
uniqueness. For each ambient pressure $P_{\rm ext}$ the obtained 
solution is a crosscut profile at $z={\rm const}$. Piling of the 
different crosscuts is a solution for an outflow in which one may 
neglect the derivatives in $z$ in comparison with the derivatives in 
$r_{\perp}$ in the two-dimensional Grad-Shafranov and Bernoulli 
equations. This can be done for highly collimated, at least 
as a parabola, outflows~\citep{NBKZ15}.

We apply here cylindrical approximation to a conical jet with
small opening angle. As $v_{\varphi}\ll v_z$ and $v_r\ll v_z$, one
has to consider all components of velocity. We assume that jet has
conical form with half opening angle $\theta \approx 2.5^{\circ}$.
It gives
\begin{eqnarray}
u_{z}(r_{\perp}) & = & u_{p} 
\cos\left({\frac{r_{\perp}}{d_{\rm jet}}\theta}\right)_,\\
u_{r}(r_{\perp}) & = & u_{p} 
\sin\left({\frac{r_{\perp}}{d_{\rm jet}}\theta}\right).
\end{eqnarray}
These formulae are to be valid also in the case of parabolic jet where 
$\theta$ is an angle between tangent and $z$-axis at the crosscut.
An important formula for calculations of the Doppler boosting factor
is for the angle between velocity of plasma parcel and
the line-of-sight $\chi$. Introducing Cartesian coordinate system 
such that $\varphi =0$ corresponds to $x$-direction and $\varphi=\pi/2$ 
corresponds to the $y$-direction, one obtains for line-of-sight 
vector
\begin{equation}
{\bf e} = (\sin{\alpha},0,\cos{\alpha}),
\end{equation}
where $\alpha$ is an angle between the jet axis $z$ and the line-of-sight. 
Hence,
\begin{equation}
\cos{\chi}=\frac{(v_{\varphi} \sin{\varphi}+v_r \cos{\varphi})\sin{\alpha}
+v_z \cos{\alpha}}
{\sqrt{v_p^2+v_{\varphi}^2}}.   
\end{equation} 

\subsection{Internal structure of relativistic jets}

\begin{figure}
    \centering
    \includegraphics[width=1\linewidth]{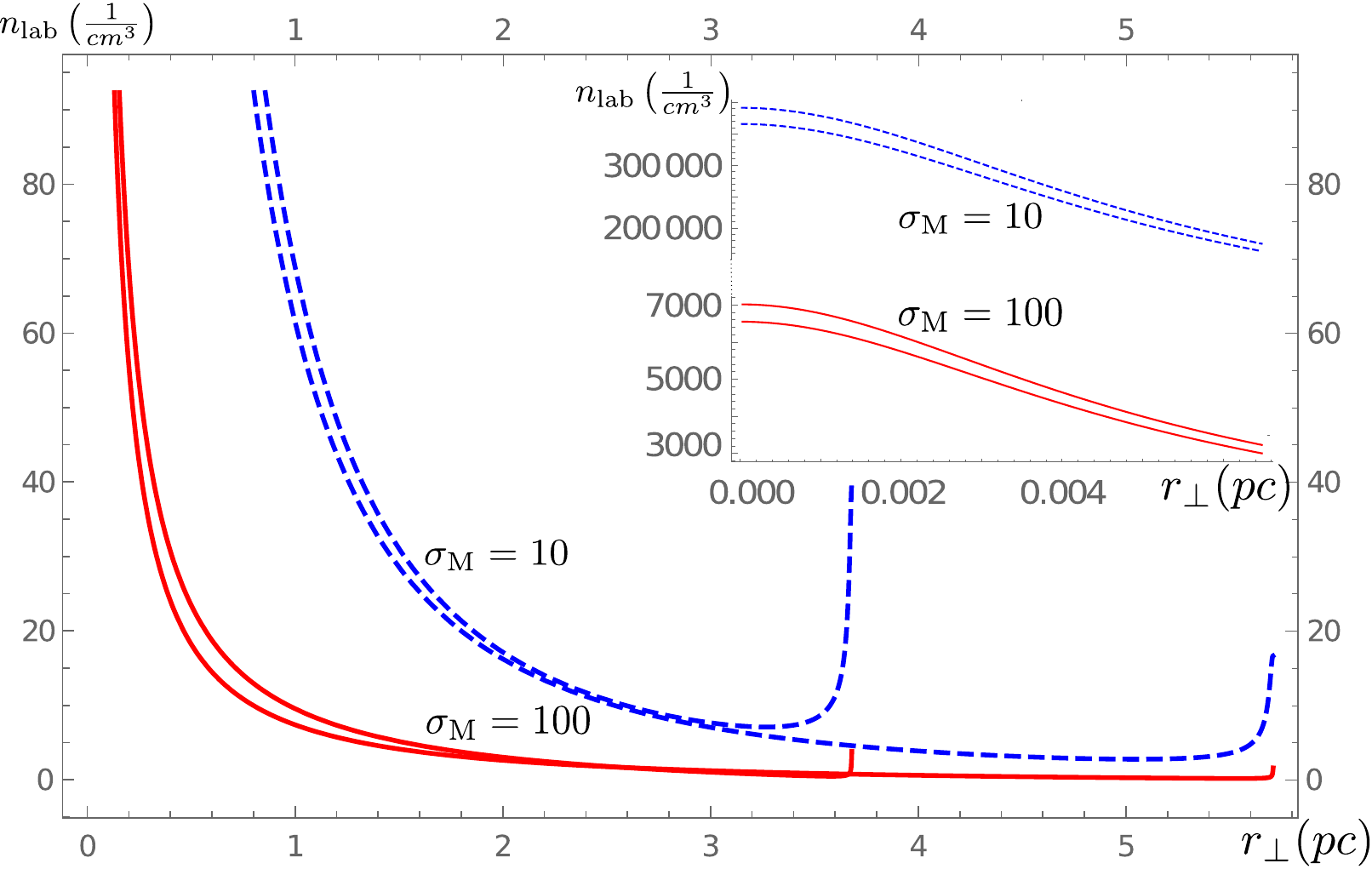}
    \caption{Number density profiles at two different cross-section of the jet in the
     laboratory frame. Red solid lines correspond to magnetisation parameter 
    $\sigma_{\rm M}=100$, and dashed blue lines to $\sigma_{\rm M}=10$. The 
     region of high densities at the axis of the jet is zoomed in.}
    \label{density}
\end{figure}


\begin{figure}
    \centering
    \includegraphics[width=1\linewidth]{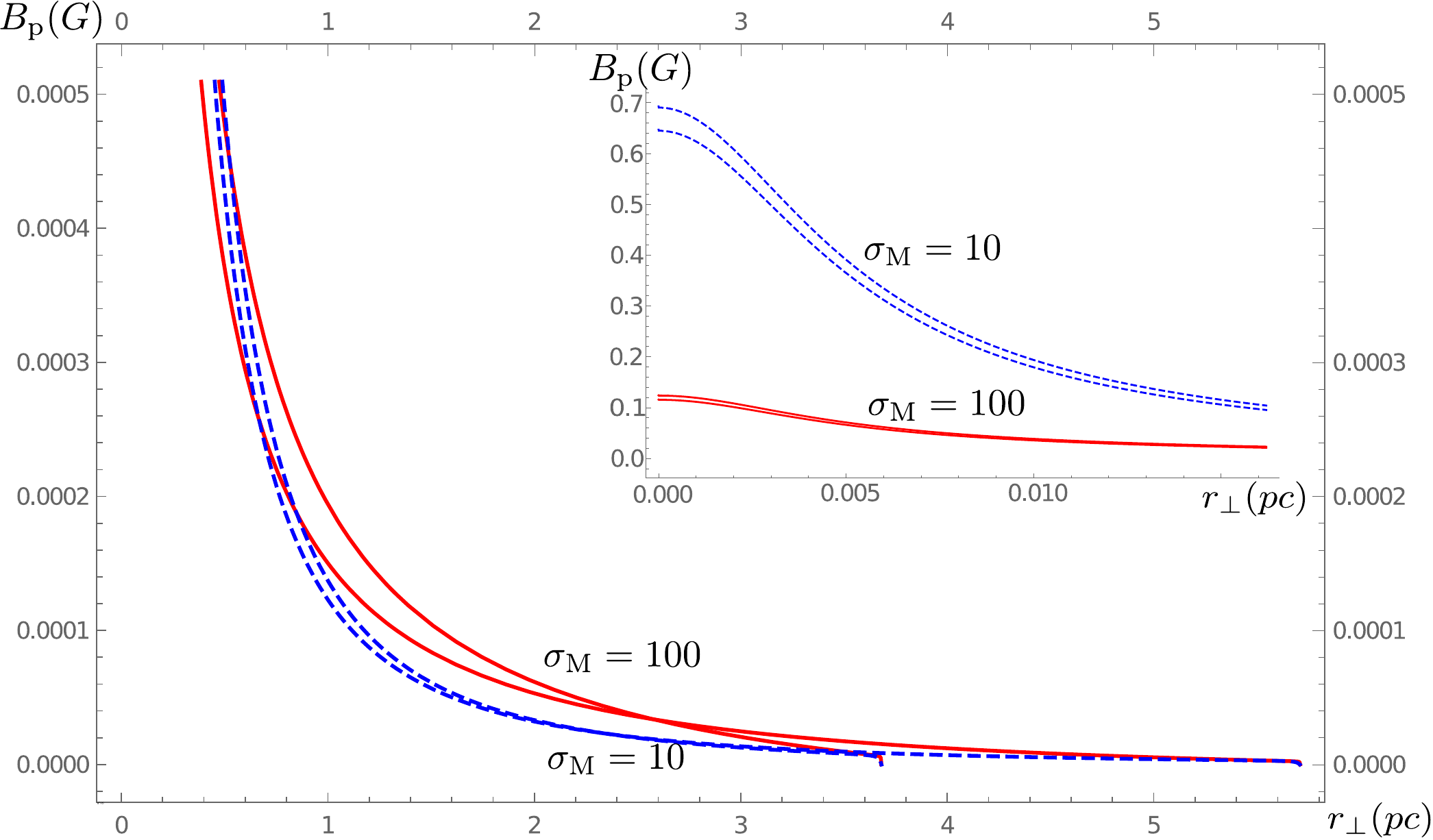}
    \caption{Poloidal magnetic field at two different cross-sections of the jet. 
    Red solid lines correspond to magnetisation parameter $\sigma_{\rm M} = 100$,
    and dashed blue lines to $\sigma_{\rm M} = 10$. The central region  is resolved.}
    \label{Bp}
\end{figure}

\begin{figure}
    \centering
    \includegraphics[width=1\linewidth]{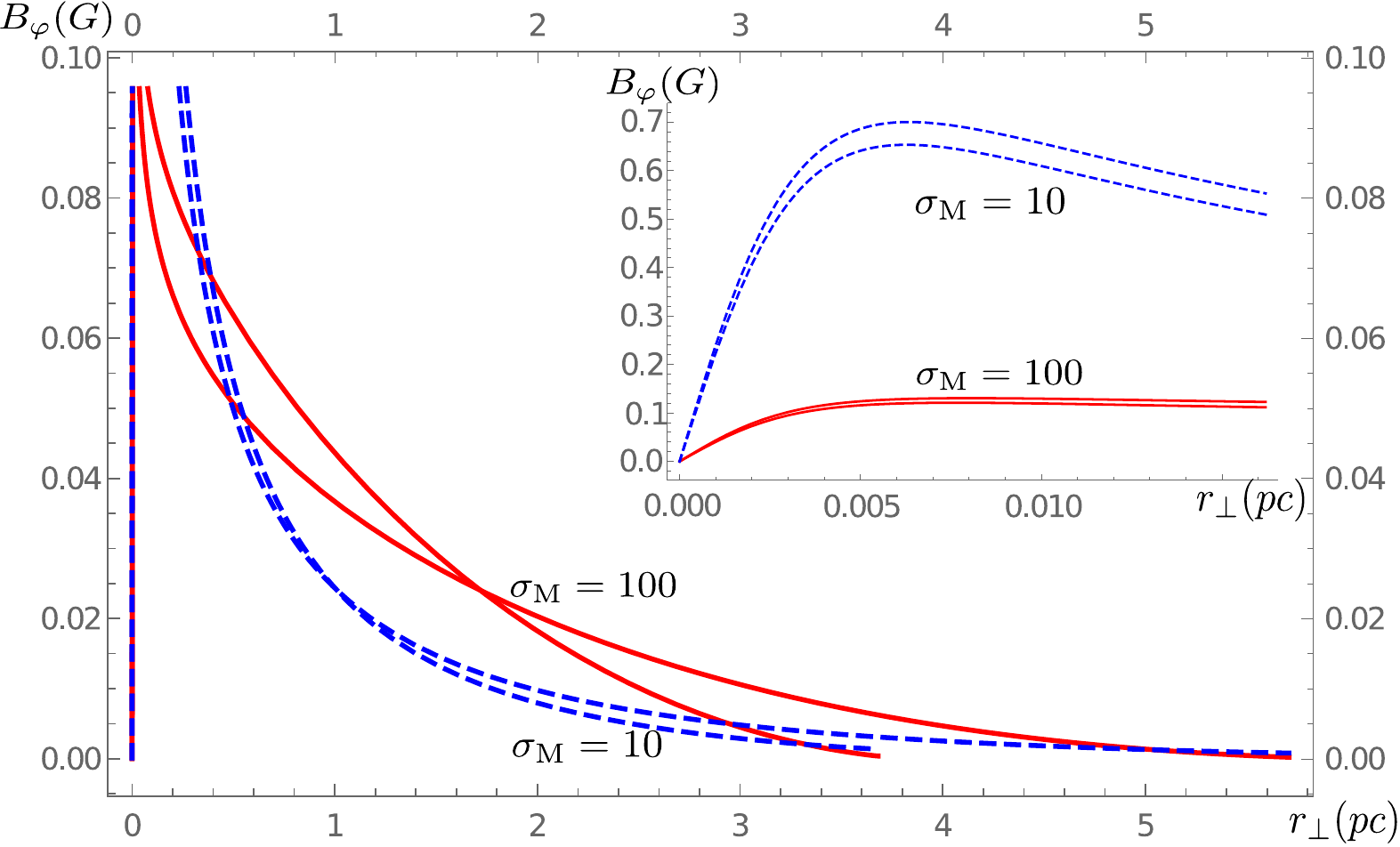}
    \caption{Toroidal component of magnetic fields at two different cross-sections 
    of the jet. Red solid lines correspond to magnetisation parameter $\sigma_{\rm M} = 100$,
    and dashed blue ones to $\sigma_{\rm M} = 10$. The central region  is resolved.}
    \label{Bphi}
\end{figure}

\begin{figure}
    \centering
    \includegraphics[width=1\linewidth]{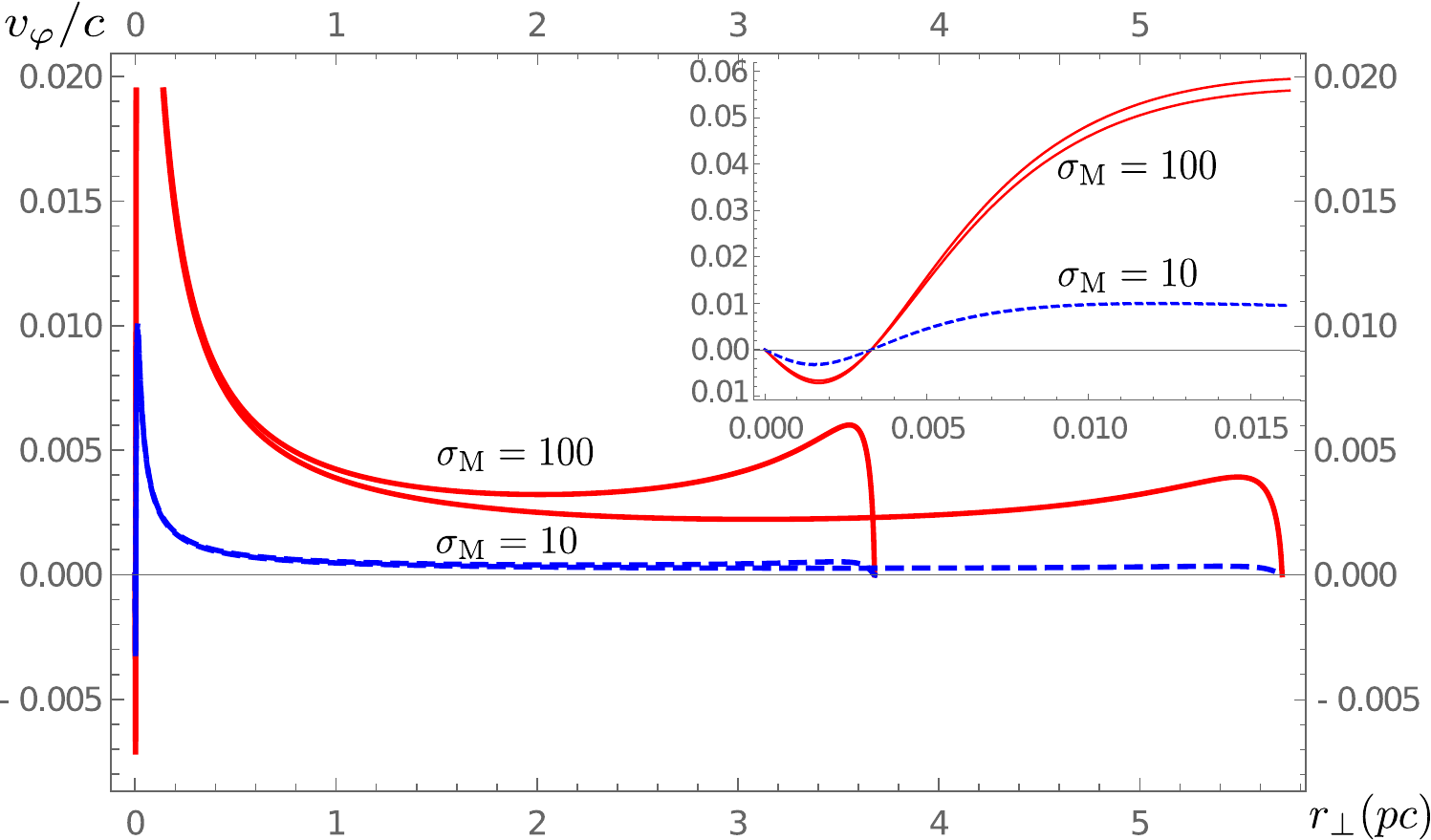}
    \caption{Toroidal components of mass velocities at two different cross-sections 
    of the jet. Plasma changes its rotational direction in the innermost region. 
    Red solid lines correspond to magnetisation parameter $\sigma_{\rm M} = 100$,
    and dashed blue lines to $\sigma_{\rm M} = 10$. The central region of peak velocity is resolved.}
    \label{vphi}
\end{figure}

\begin{figure}
    \centering
    \includegraphics[width=1\linewidth]{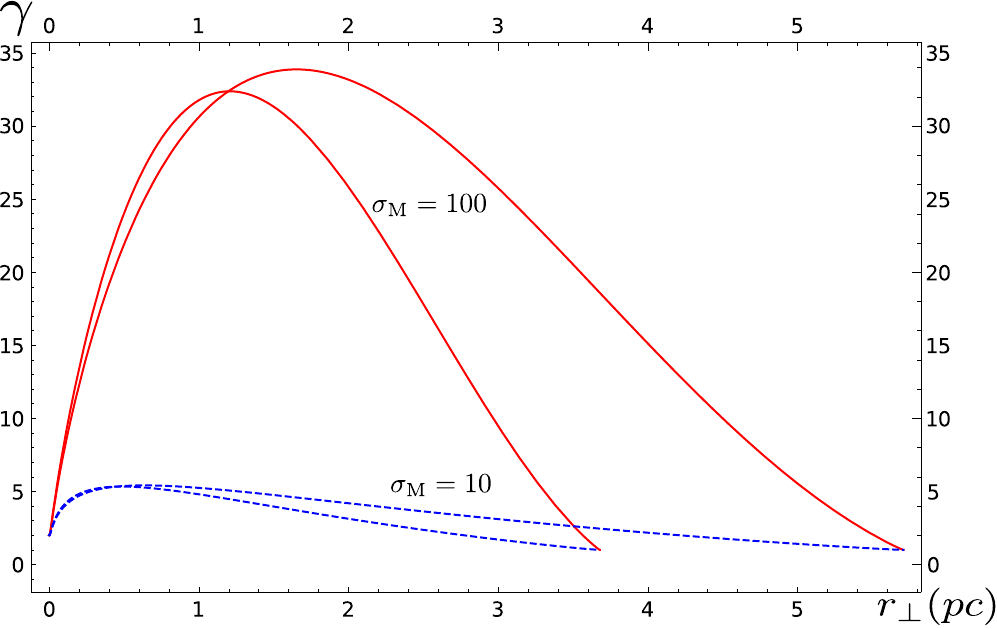}
    \caption{Profiles of the Lorentz factors at two different cross-sections of the jet.
    Red solid lines correspond to magnetisation $\sigma_{\rm M}=100$,
    and dashed blue lines correspond to $\sigma_{\rm M}=10$. Lorentz factor at the axis
    $\gamma_{\rm in}=2$.}
    \label{gamma}
\end{figure}

\begin{figure*}
\begin{minipage}[h]{0.47\linewidth}
\center{\includegraphics[width=1\linewidth]{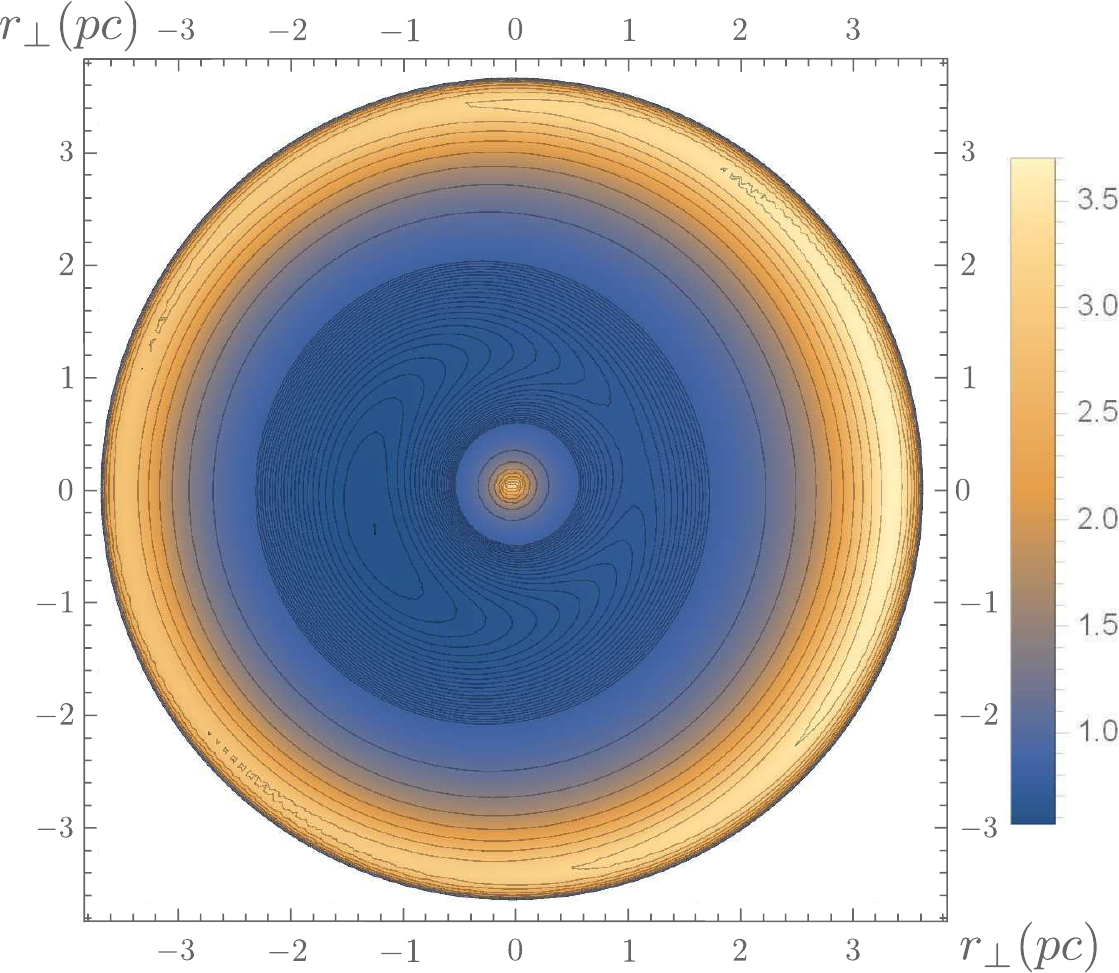}} a) \\
\end{minipage}
\hfill
\begin{minipage}[h]{0.47\linewidth}
\center{\includegraphics[width=1\linewidth]{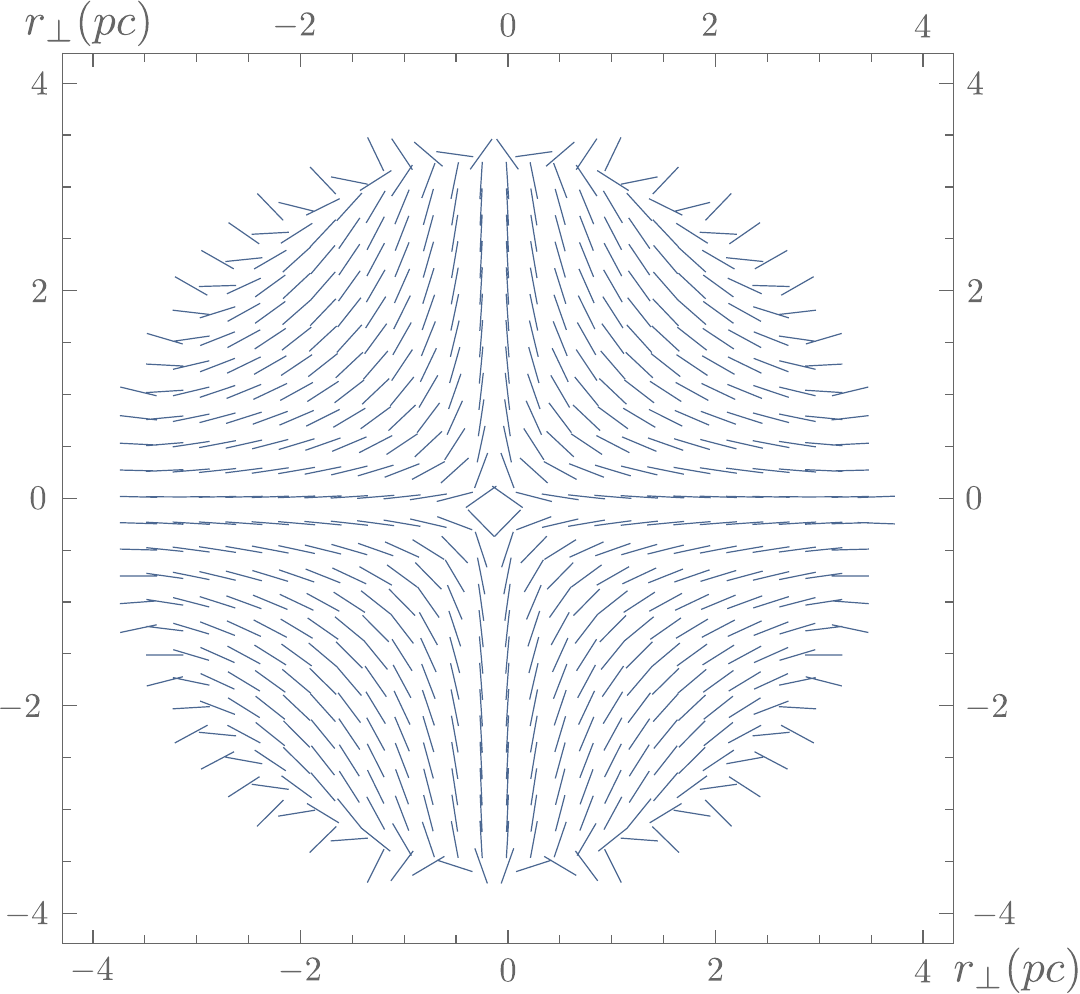}} \\b)
\end{minipage}
\vfill
\begin{minipage}[h]{0.47\linewidth}
\center{\includegraphics[width=1\linewidth]{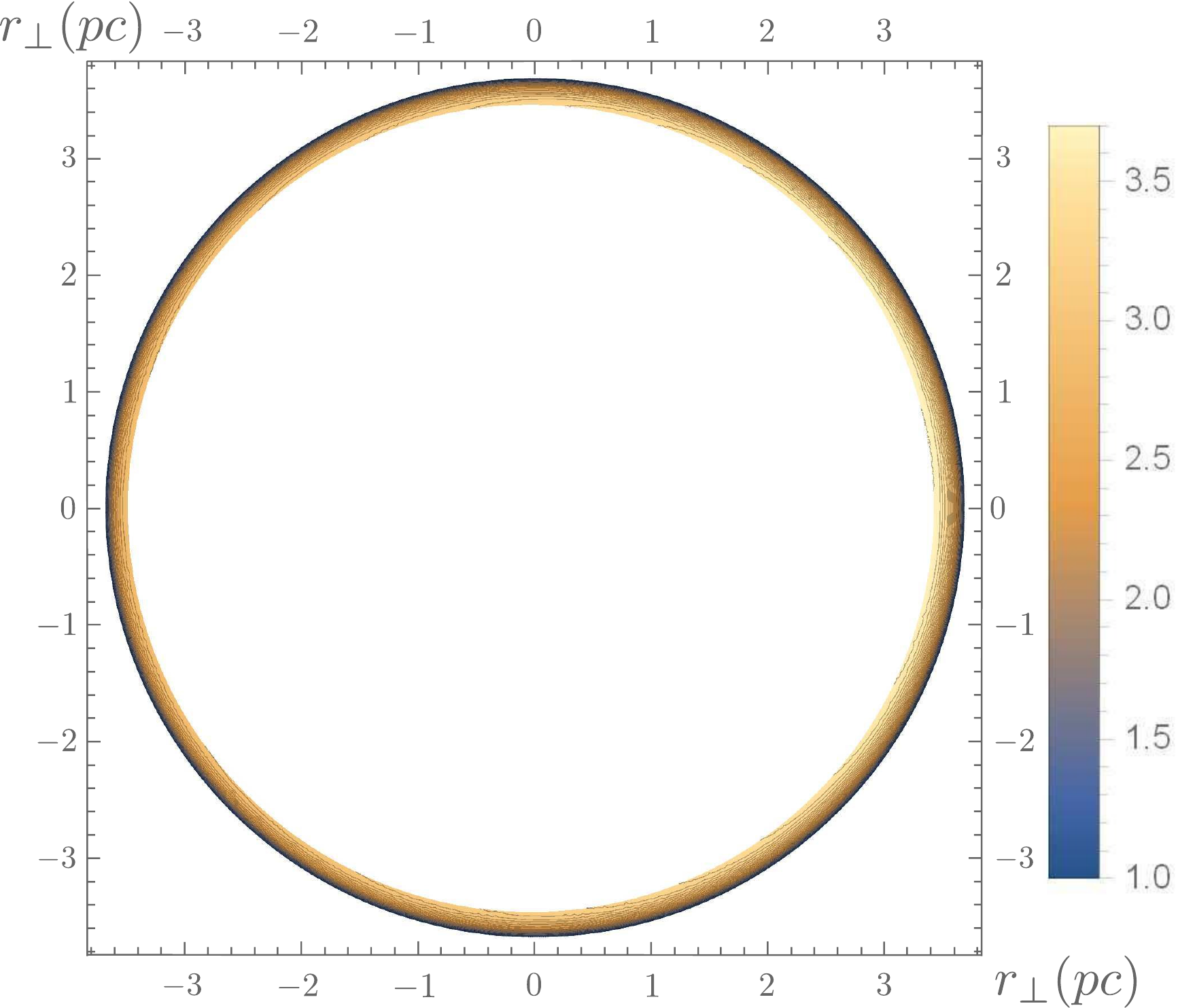}} c) \\
\end{minipage}
\hfill
\begin{minipage}[h]{0.47\linewidth}
\center{\includegraphics[width=1\linewidth]{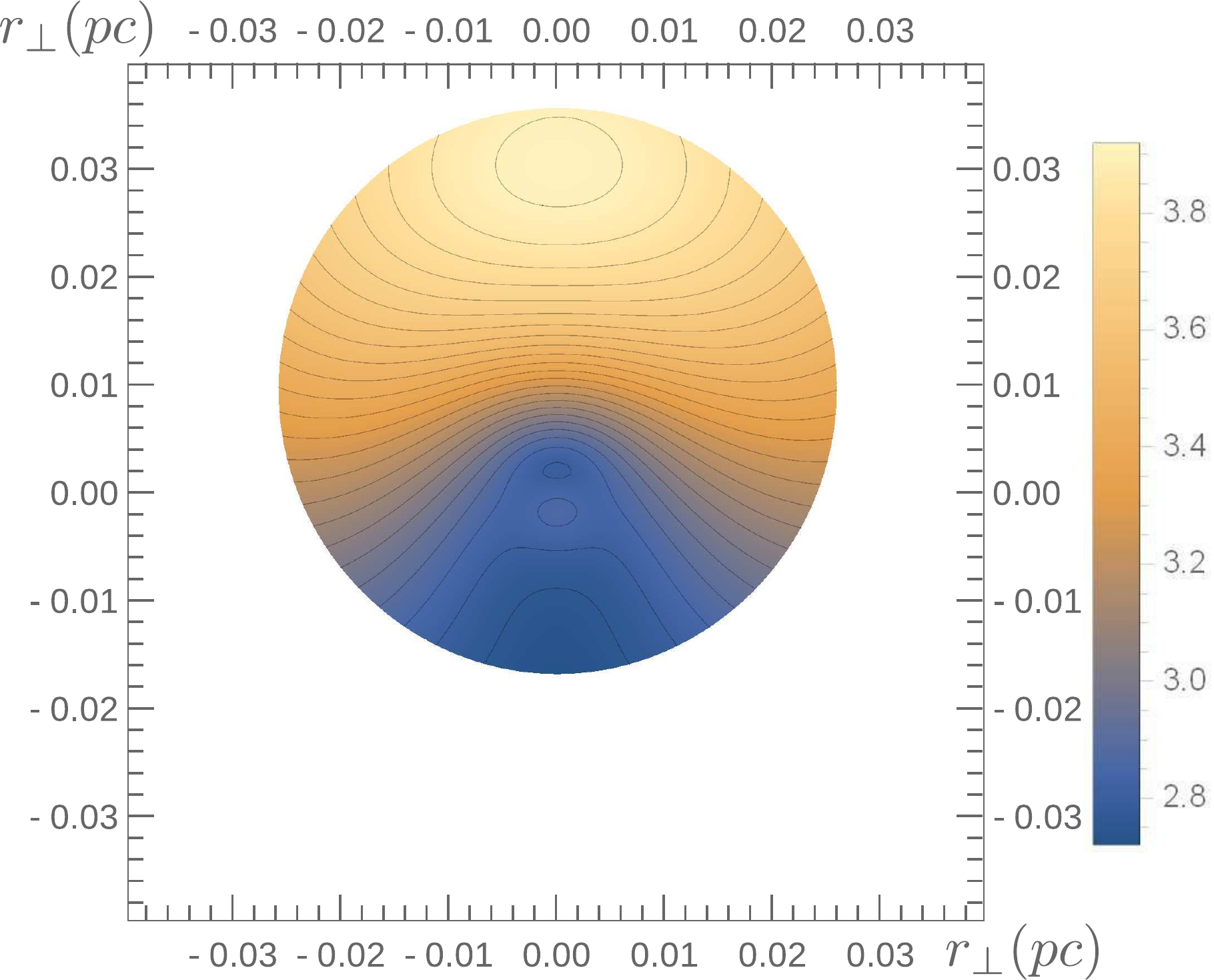}} d) \\
\end{minipage}
\caption{ The distribution of the Doppler factor on the cross section 
of the jet for inclination angle  $\alpha = 18^{\circ}$ for parameters of M87 
and $\sigma_{\rm M}=100$, $\gamma_{\rm in}=2$ as would be seen by an observer 
looking down on the jet from the above and from the right at an angle
$\alpha = 18^{\circ}$ with respect to the jet axis:
a) Map of Doppler boosting factor as 
a whole. Contour lines are drawn with a step 
0.01 for \mbox{$\delta \in (0, 0.8)$} and 0.3 for 
\mbox{$\delta \in (0.8, 4)$.} b) Map of the directions of electric 
vector of linearly polarised radio emission. c) Map of radiation which is 
not depressed by relativistic beaming effect. d) Zoom in on the 
central region of the map (c) which is not resolved on plot (c).} 
\label{Doppler18}
\end{figure*}

\begin{figure*}
\begin{minipage}[h]{0.47\linewidth}
\center{\includegraphics[width=1\linewidth]{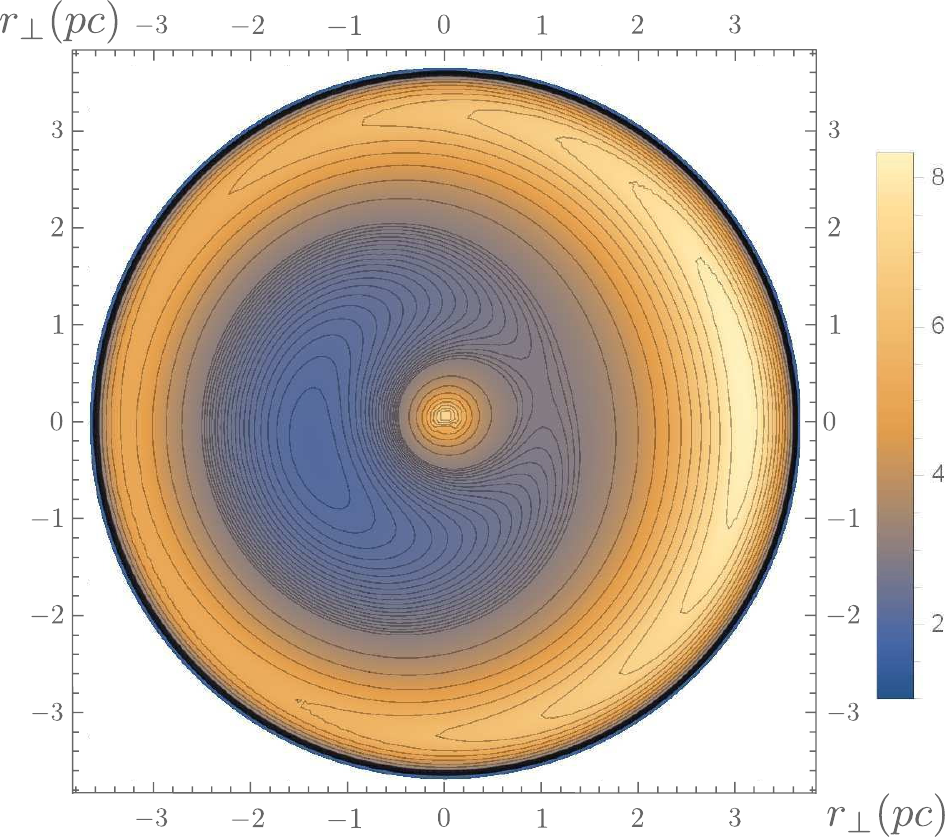}} a) \\
\end{minipage}
\hfill
\begin{minipage}[h]{0.47\linewidth}
\center{\includegraphics[width=1\linewidth]{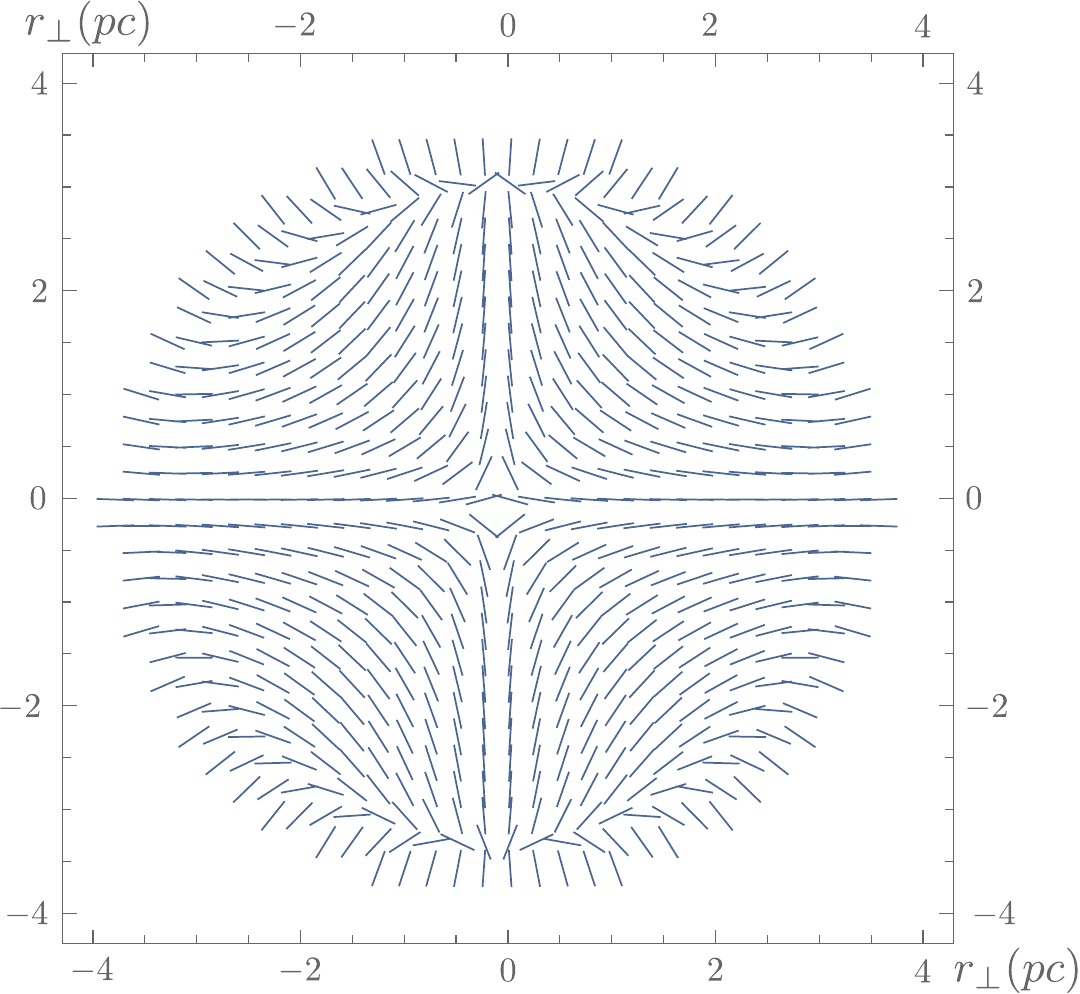}} \\b)
\end{minipage}
\vfill
\begin{minipage}[h]{0.47\linewidth}
\center{\includegraphics[width=1\linewidth]{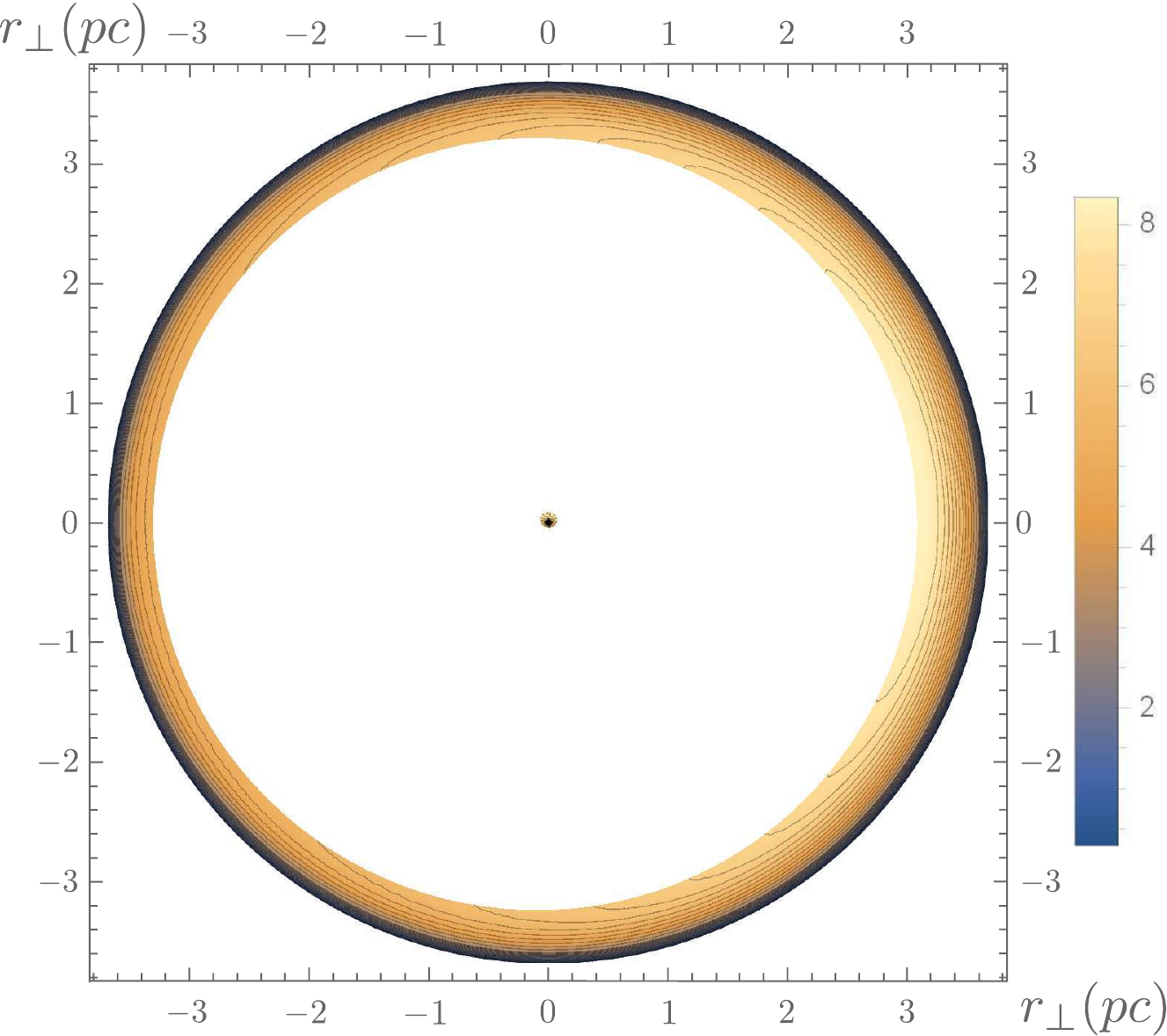}} c) \\
\end{minipage}
\hfill
\begin{minipage}[h]{0.47\linewidth}
\center{\includegraphics[width=1\linewidth]{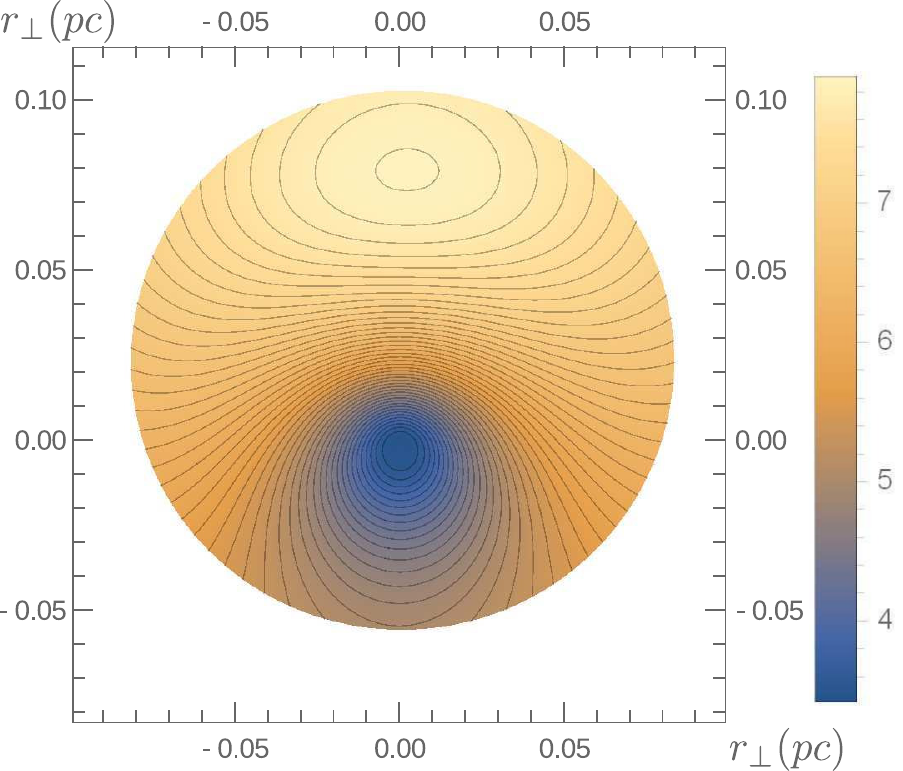}} d) \\
\end{minipage}
\caption{The same as at Fig.~\ref{Doppler18} but for inclination angle  $\alpha
=9^{\circ}$.  Magnetisation parameter $\sigma_{\rm M}=100$. Contour 
lines are drawn with step 0.05 for \mbox{$\delta \in (0, 3)$} and 
0.3 for \mbox{$\delta \in (3, 8)$.}}  
\label{Doppler9}
\end{figure*}

As was already stressed, for given five ''integrals of motion'' 
the solution at each cross-section is fully determined by the 
value of the ambient pressure $P_{\rm ext}$. It concerns the 
jet width $d_{\rm jet}$ as well. For this reason in what follows 
we use the jet thickness $d_{\rm jet}$ as a main parameter as 
it can be directly determined from observations.

On the other hand, as was shown by~\citet{BCKN-17}, our choice of 
''integrals of motion'' allows us to express them through only one 
dimensionless quantity:
\begin{equation}
\sigma_{\rm M} = \frac{\Omega_{0}^2 \Psi_{\rm tot}}{8 \pi^2 \mu\eta c^2},
\label{sigmaM}
\end{equation}
i.e., through the Michel magnetisation parameter, which is the parameter of a
''central engine'' (it gives the maximum bulk Lorentz factor of the outflow). 
In addition, we use parameters of M87 black hole for calculations below. 
Another assumption is the value of 
the regular magnetic field nearby the event horizon $B=10^4$ Gs. It is also assumed 
that magnetisation parameter $\sigma_{\rm M} = 10-100$. This choice is reasonable 
for AGNs (see, e.g., \citealt{NBKZ15}). 


 
The profiles of magnetic field components, velocities of particles, number density, 
and Lorentz factor of plasma are presented in Figs.~\ref{density}--\ref{gamma}
for two different width of the jet $d_{\rm jet}$ and two different 
values of magnetisation parameter $\sigma_{\rm M}$.
As we show in Fig.~\ref{density}, our choice of integrals of motion results 
in fast decrease of the number density in laboratory frame $n_{\rm e}=n \gamma$ with
the distance from the rotation axis, where $n$ is found via (\ref{M2}). As was mentioned 
above, number density is determined by magnetisation parameter $\sigma_{\rm M}$. 
As we see, number density is larger for smaller $\sigma_{\rm M}$. 
The dramatic growth of the density at the jet boundary is just 
dictated by pressure balance inside the thin boundary layer~\citep{BCKN-17} 
\begin{equation}
P + \frac{B^2}{8 \pi} = {\rm const}   
\end{equation}
and by vanishing magnetic field outside the jet. Inside the jet 
the ``cavity'' is supported by large magnetic field pressure 
and by centrifugal force. 

\section{Results}
\label{results}

Further, on Figs.~\ref{Bp} and \ref{Bphi} we show the structure of magnetic
field inside the jet. As we see, the magnetic field also forms the central core 
and then drops towards the jet boundary. In the narrow central part the toroidal 
component growth linearly as $B_{\varphi} \propto I/r$ and $I=\pi j_{\rm p} r^2$. 
Here $j_{\rm p}$ is the current density. Outside the light cylinder $R_{\rm L}$ 
toroidal magnetic field prevails up to the very edge $B_{\varphi} \gg B_{\rm p}$. 
Since the jet radius in quasi-cylindrical part is 
$d_{\rm jet} \approx 10^2-10^4 R_{\rm L}$,
the relative size of region $B_{\varphi}<B_{\rm p}$ is extremely small. 

Besides, on Fig.~\ref{vphi} we show the toroidal components of
hydrodynamical velocity $v_{\varphi}$.
The differences in the magnitude are attributed to the particle-to-magnetic flux 
ratio which is larger for smaller magnetisation parameter (\ref{sigmaM}).
In any way, maximum value for toroidal velocity $v_{\varphi}$ cannot 
exceed a few tenths of speed of light $c$. In this sense our predictions 
do not contradict observational data~\citep{Mertens}. On the other hand, 
as one can see directly from (\ref{p35}), toroidal velocity $v_{\varphi}$ 
can hardly be determined theoretically. The point is that two terms 
in the numerator have the same order of magnitude, the first one being 
related to the sliding along the magnetic surface, and the second one being 
related to the angular momentum. E.g., for monopole magnetic field (and for slow 
rotation) the toroidal velocity vanishes~\citep{Bogovalov92, BO}. For 
this reason it is not surprising that $v_{\varphi}$ can change sign.

\begin{figure*}
\begin{minipage}[h]{0.47\linewidth}
\center{\includegraphics[width=1\linewidth]{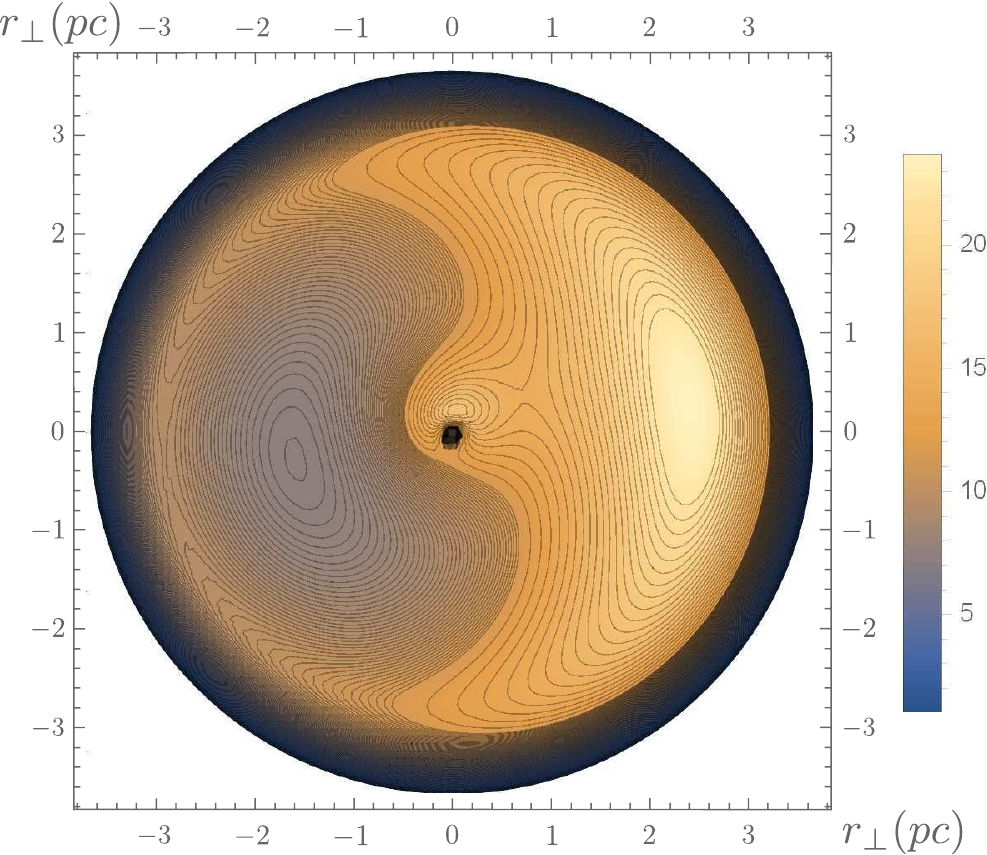}} a) \\
\end{minipage}
\hfill
\begin{minipage}[h]{0.47\linewidth}
\center{\includegraphics[width=1\linewidth]{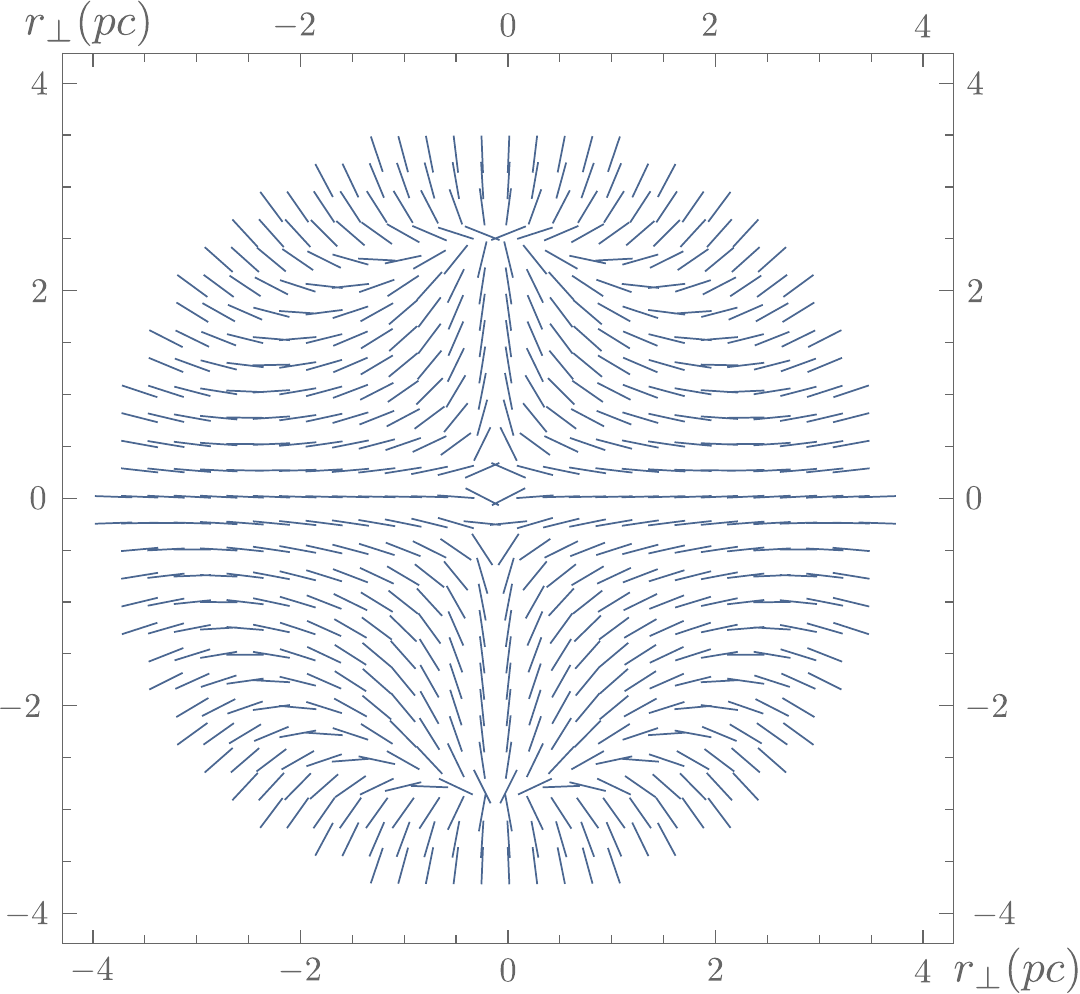}} \\b)
\end{minipage}
\vfill
\begin{minipage}[h]{0.47\linewidth}
\center{\includegraphics[width=1\linewidth]{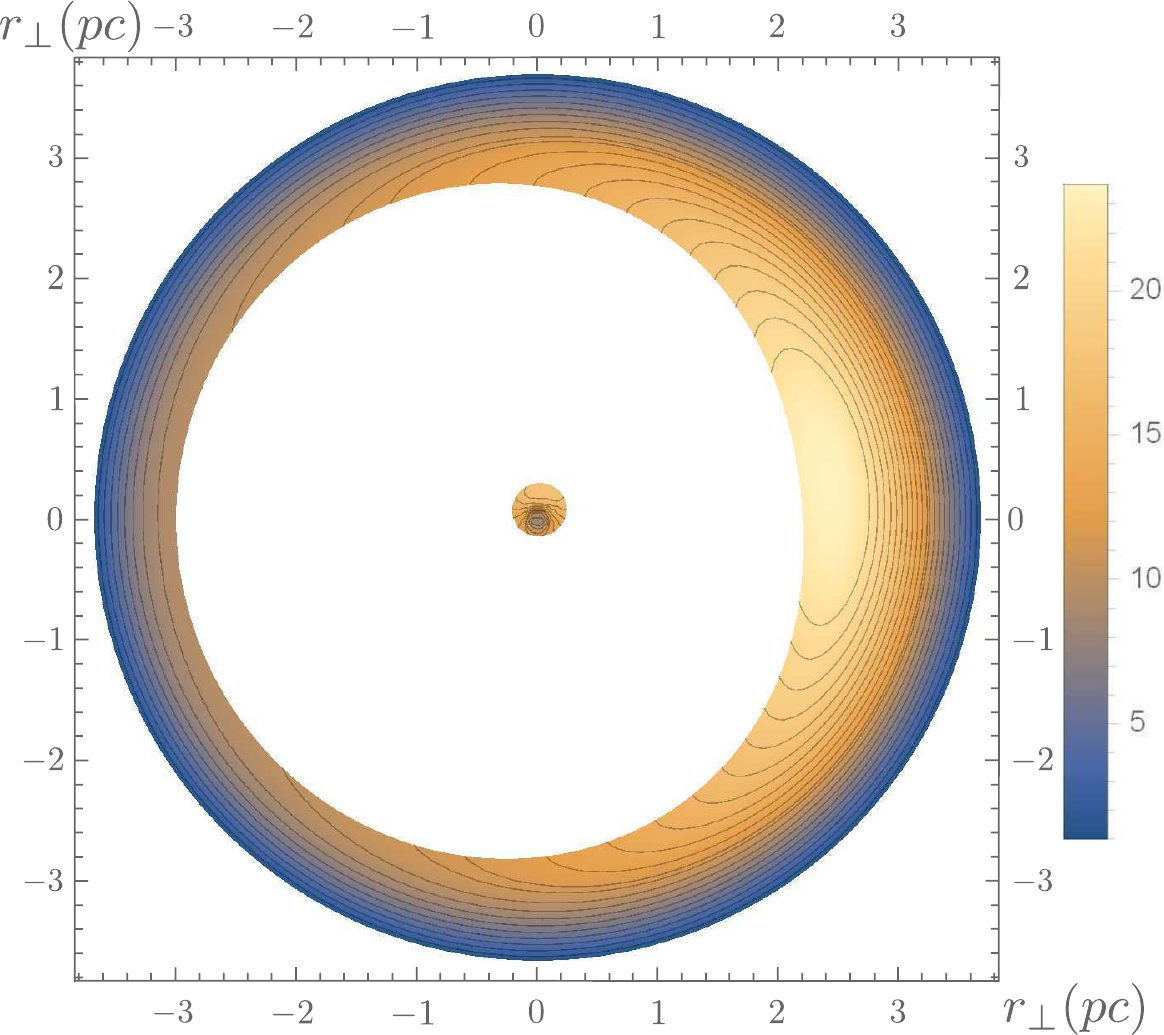}} c) \\
\end{minipage}
\hfill
\begin{minipage}[h]{0.47\linewidth}
\center{\includegraphics[width=1\linewidth]{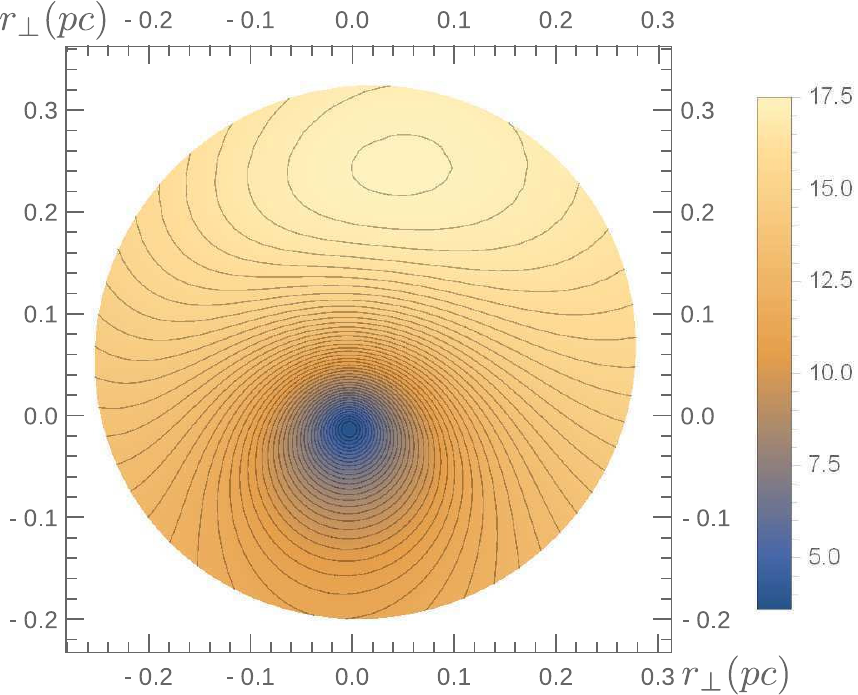}} d) \\
\end{minipage}
\caption{
The same as at Fig.~\ref{Doppler18} but for inclination angle  $\alpha
=4^{\circ}$.  Magnetisation parameter $\sigma_{\rm M}=100$. Contour 
lines are drawn with step 0.1 for \mbox{$\delta \in (0, 11.5)$} and 
0.5 for \mbox{$\delta \in (11.5, 20)$.}}
\label{Doppler4}
\end{figure*}

\begin{figure*}
\begin{minipage}[h]{0.47\linewidth}
\center{\includegraphics[width=1\linewidth]{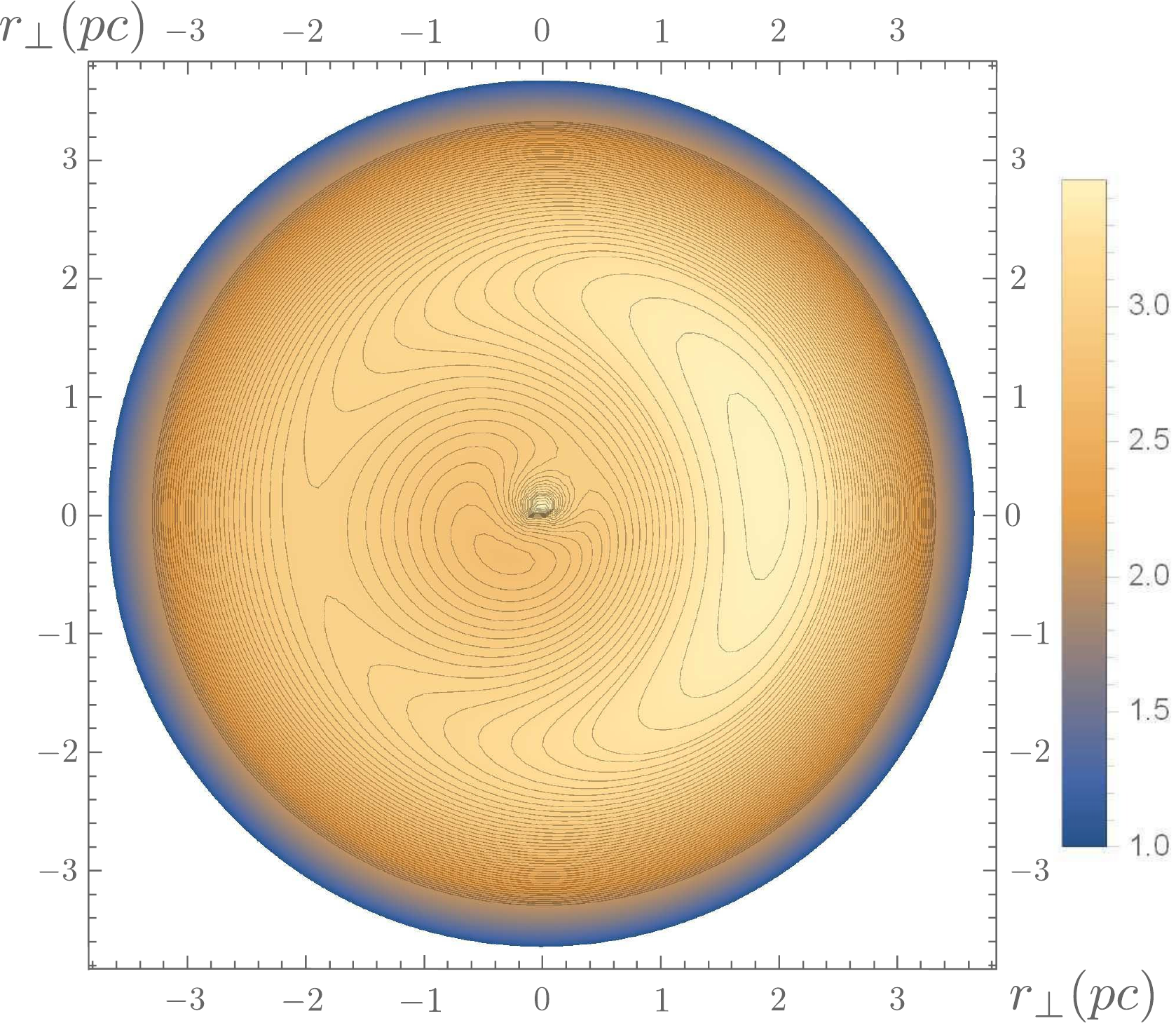}} a) \\
\end{minipage}
\hfill
\begin{minipage}[h]{0.47\linewidth}
\center{\includegraphics[width=1\linewidth]{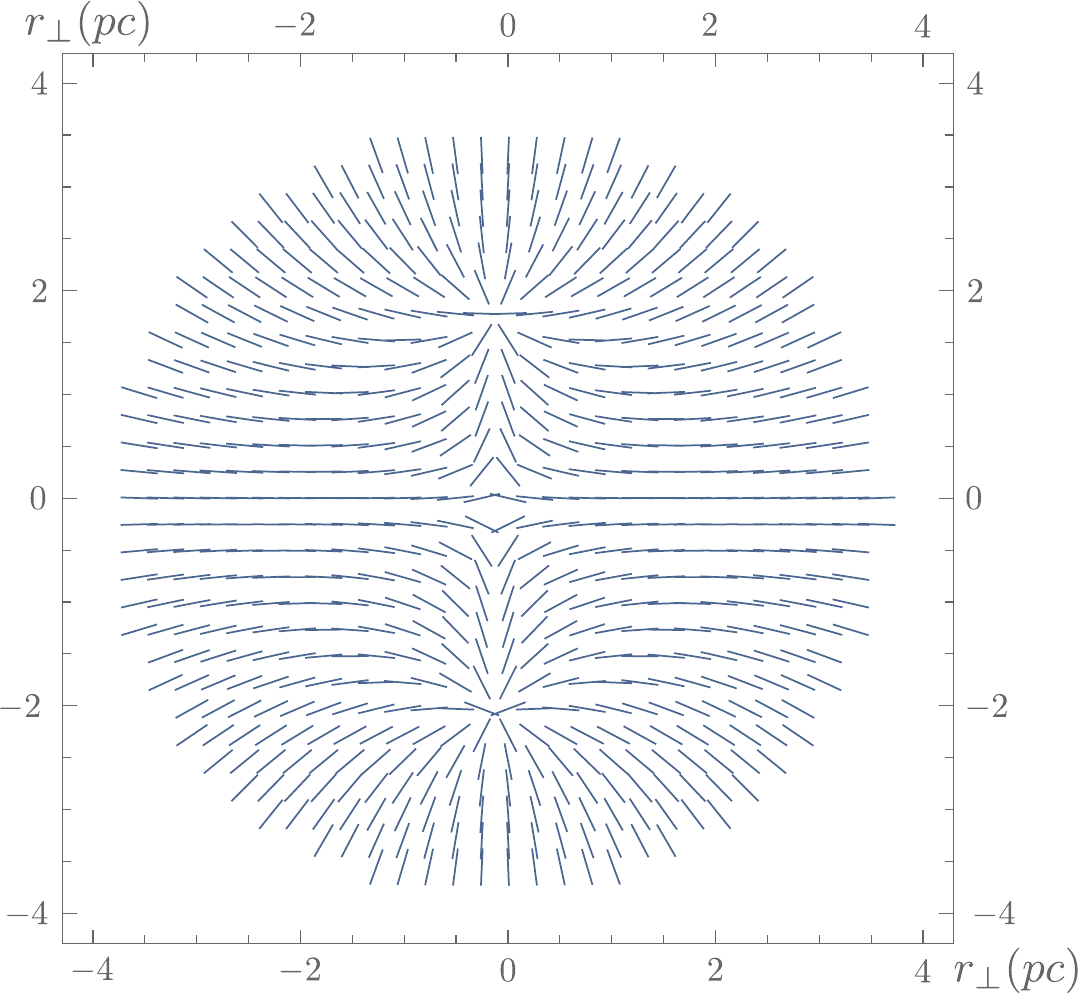}} \\b)
\end{minipage}
\vfill
\begin{minipage}[h]{0.47\linewidth}
\center{\includegraphics[width=1\linewidth]{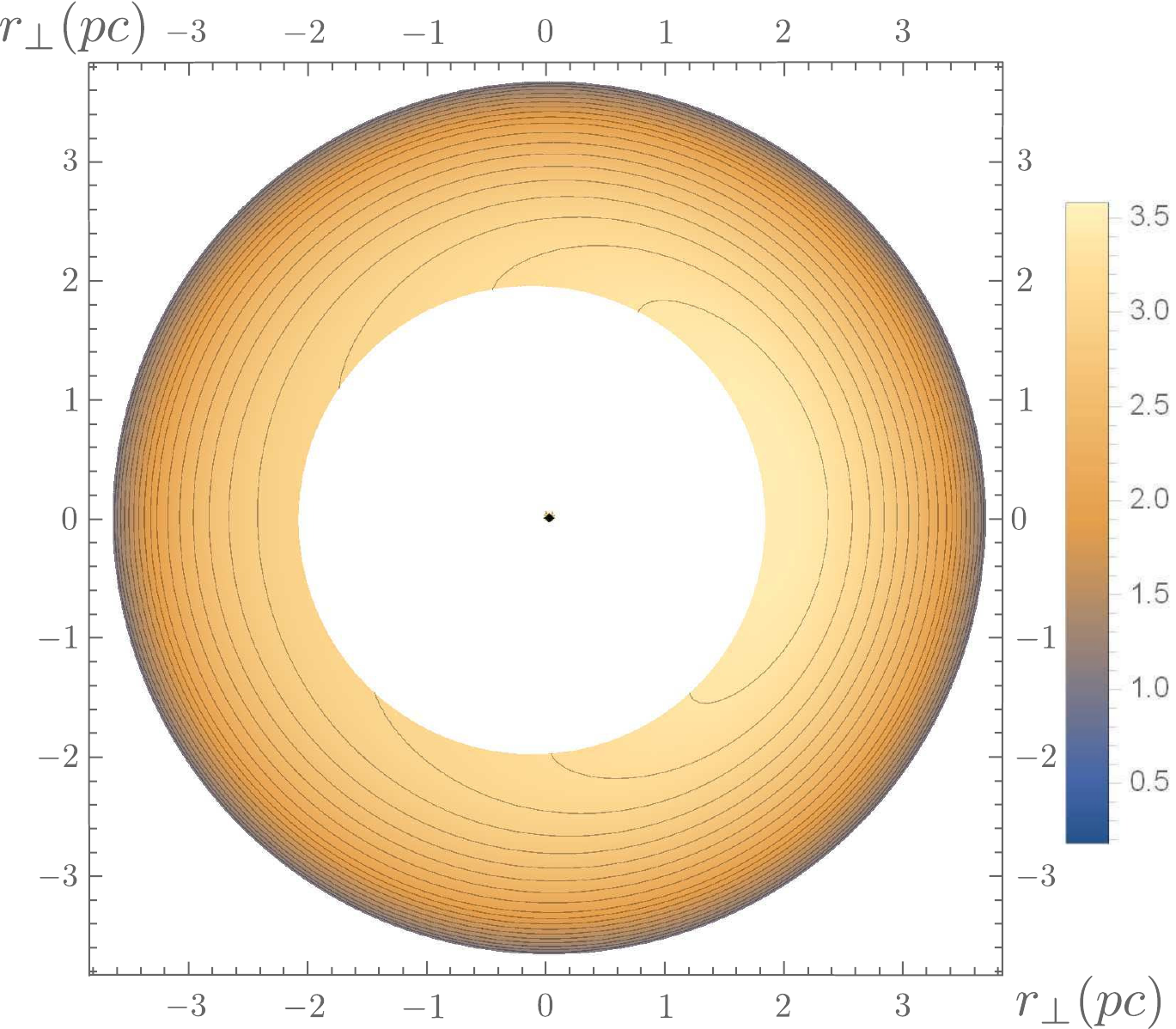}} c) \\
\end{minipage}
\hfill
\begin{minipage}[h]{0.47\linewidth}
\center{\includegraphics[width=1\linewidth]{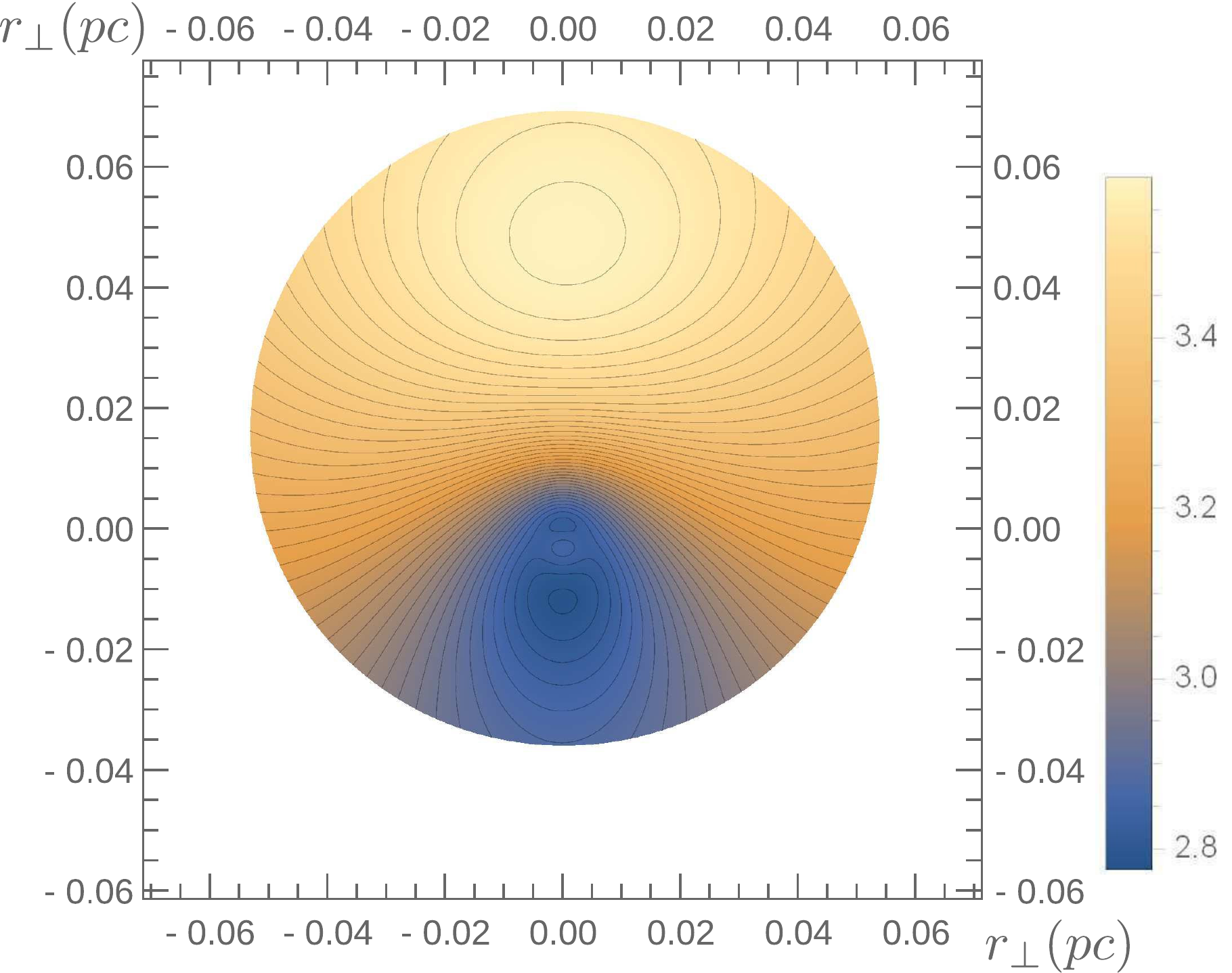}} d) \\
\end{minipage}
\caption{
The same as at Fig. \ref{Doppler18} for inclination angle  $\alpha
=18^{\circ}$ and magnetisation parameter $\sigma_{\rm M}=10$. Contour 
lines are drawn with step 0.01 for \mbox{$\delta \in (0, 2)$} and 
step 0.1 for \mbox{$\delta \in (2, 4)$.}}
\label{Doppler10}
\end{figure*}

\begin{figure*}
\begin{minipage}[h]{0.47\linewidth}
\center{\includegraphics[width=1\linewidth]{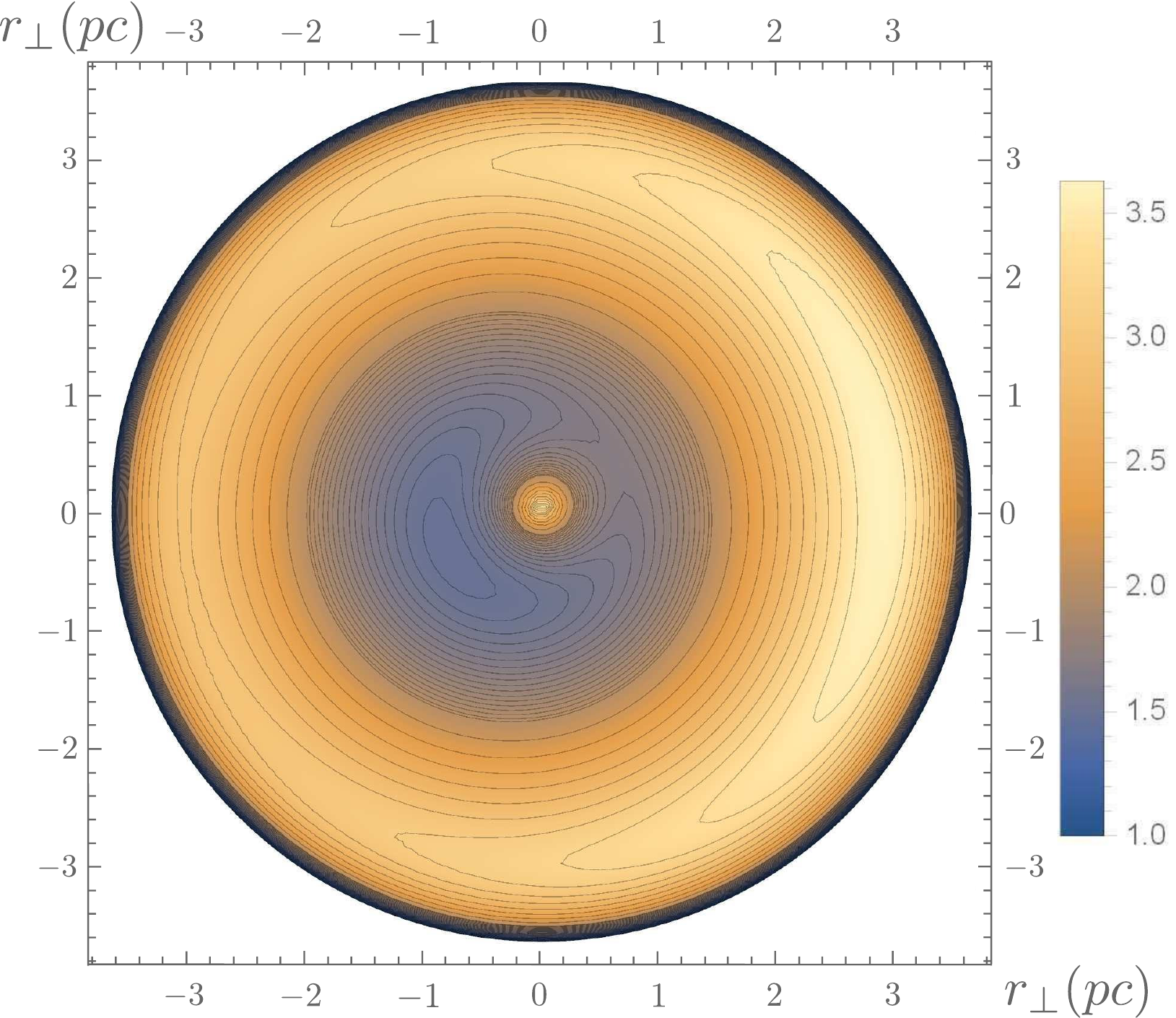}} a) \\
\end{minipage}
\hfill
\begin{minipage}[h]{0.47\linewidth}
\center{\includegraphics[width=1\linewidth]{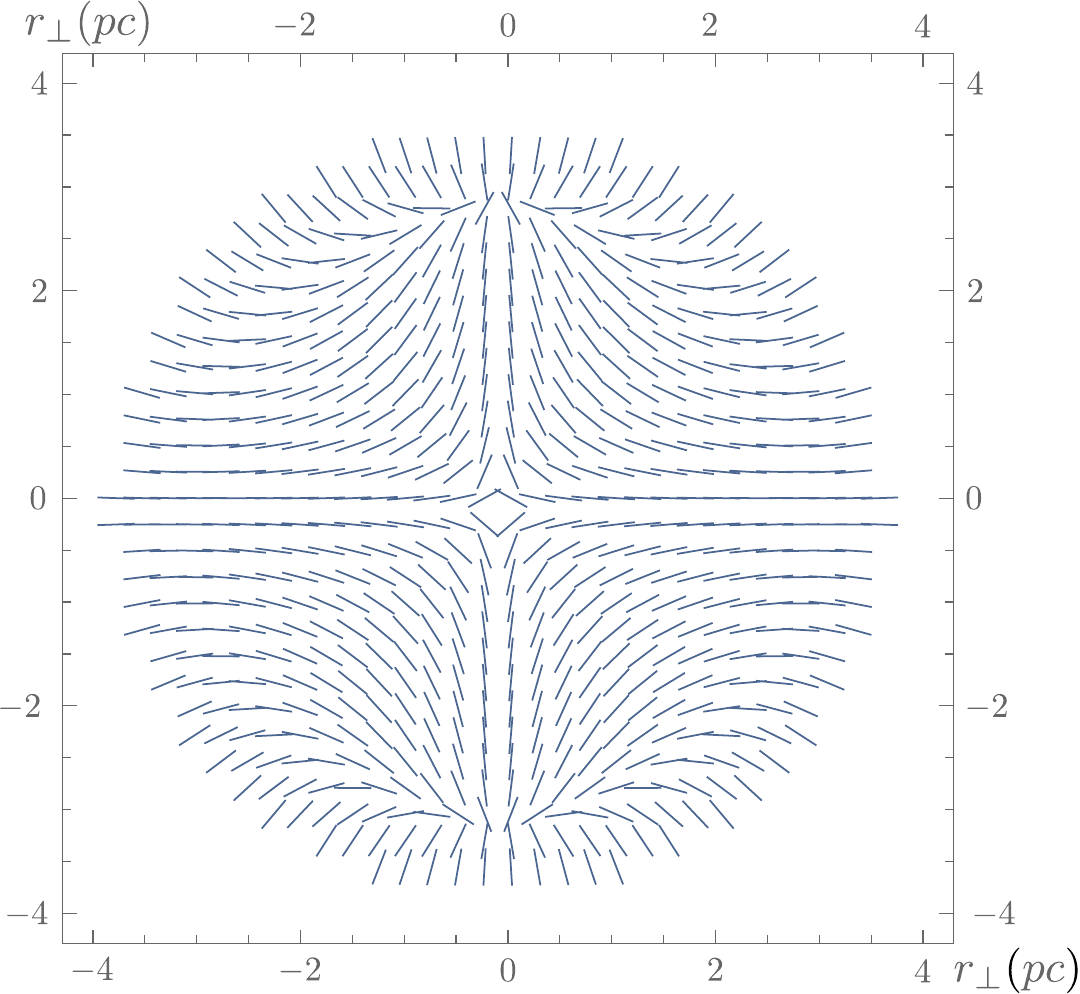}} \\b)
\end{minipage}
\vfill
\begin{minipage}[h]{0.47\linewidth}
\center{\includegraphics[width=1\linewidth]{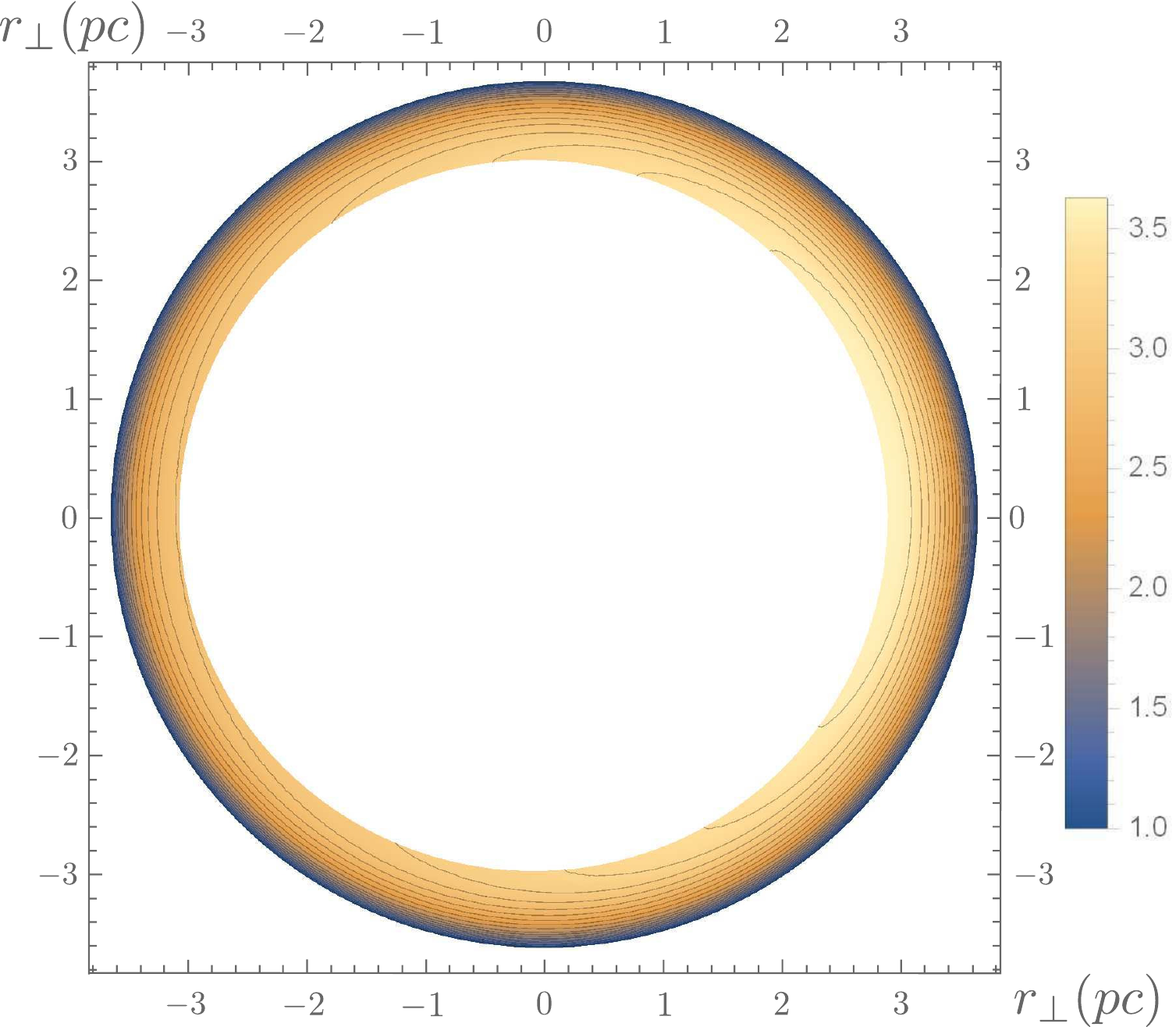}} c) \\
\end{minipage}
\hfill
\begin{minipage}[h]{0.47\linewidth}
\center{\includegraphics[width=1\linewidth]{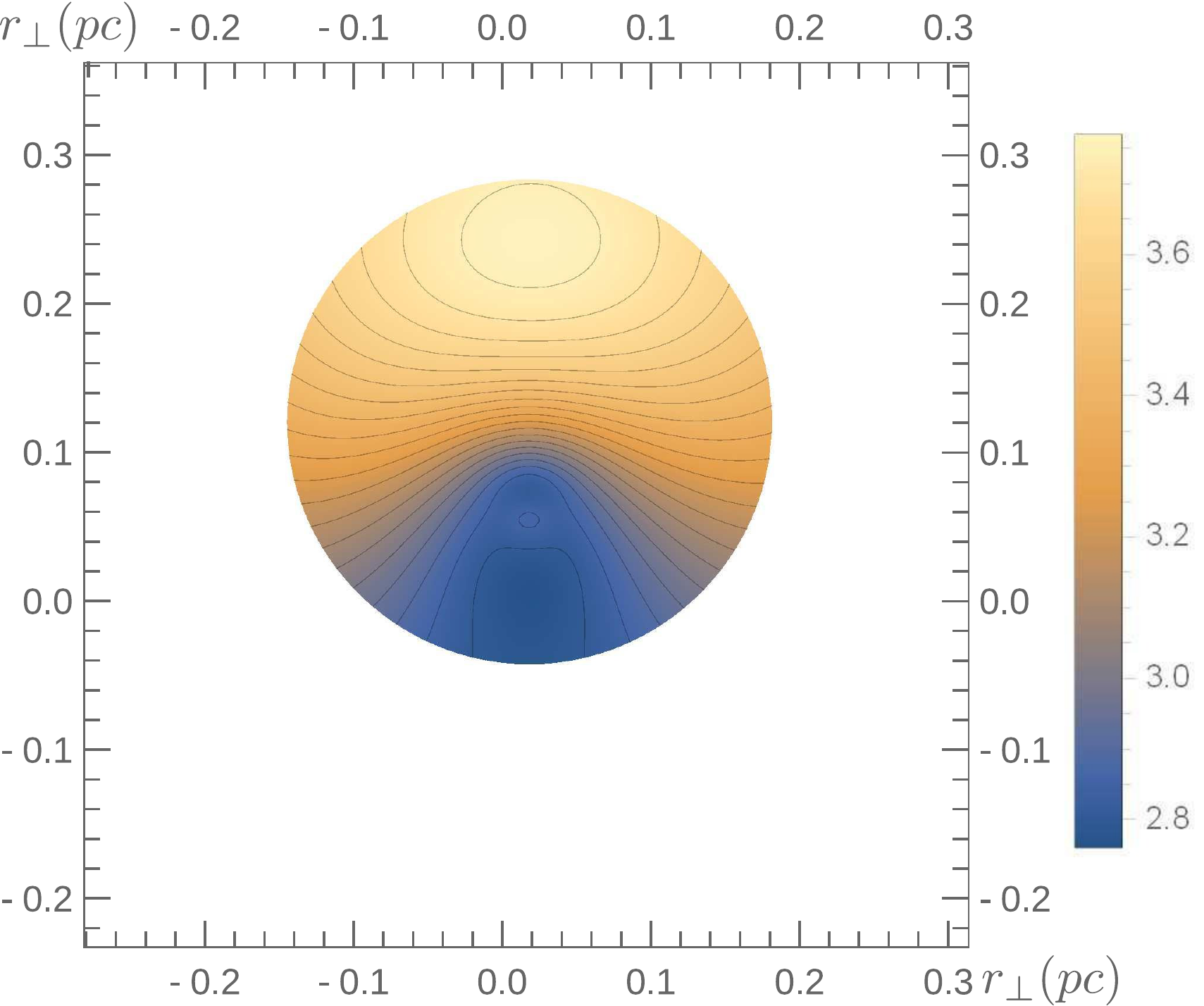}} d) \\
\end{minipage}
\caption{
The same as at Fig. \ref{Doppler18} for inclination angle  $\alpha
=18^{\circ}$ and magnetisation parameter $\sigma_{\rm M}=30$. Contour 
lines are drawn with step 0.01 for \mbox{$\delta \in (0, 0.8)$} and 
0.3 for \mbox{$\delta \in (0.8, 4)$.}}
\label{Doppler30}
\end{figure*}

Finally, as is shown on Fig.~\ref{gamma}, the magnitude of the Lorentz factor is also
determined by magnetisation parameter, its value being larger for wider jet (and,
certainly, for larger magnetisation parameter $\sigma_{\rm M}$). This fact is the
illustration of the well known dependence (\ref{gammaff}). Indeed, it can be seen that
in the central part of the jet Lorentz factor grows linearly as our choice of 
''integrals of motion'' coincides with force-free choice in the vicinity of the axis. 

It is also necessary to stress that there is no acceleration of particles along the 
rotational axis because the flux of electromagnetic energy $E \propto \Omega_{\rm F} I$ 
is equal to zero there and Lorentz-factor at the axis is chosen $\gamma_{\rm in}=2$. 
The Lorentz factor does not also change at the boundary because $I$ is zero there too. 
In contrast, the Lorenz factor intensively changes at middle radii where according to
(\ref{gammain}) the magnetisation is initially high. 

\subsection{Doppler maps}

The maps of the Doppler factor $\delta$ (\ref{delta}) are presented in 
\mbox{Figs.~\ref{Doppler18}--\ref{Doppler30}} for different inclination angles 
$\alpha$ as well as for different magnetisation parameters $\sigma_{\rm M}$. 
Comparing now the Lorentz factor distribution with the Doppler maps 
one can conclude that for large enough $\alpha$ radiation from the regions 
with highest Lorentz factor cannot be detected. 
These components may be seen in BL Lac objects only. 

Indeed, the constraints on the visible part of a jet are governed by relativistic 
beaming effect. If an angle $\chi$ between line-of-sight and velocity of plasma  
is greater than $1/\gamma$, radiation cannot be detected. As a result, as shown in
Fig.~\ref{Doppler18}c, observer can see only the regions with small enough $\gamma$,
i.e., outer parts of a jet and the jets core. E.g., for M87 (the black hole mass 
\mbox{$M=3 \times 10^9 M_{\odot}$,} $a=0.1$, distance 17 Mpc and 
inclination angle $\alpha = 18^{\circ}$) the angular size of central bright core 
is $\sim 10^{-1}$ mas, while the peak angular resolution of VLBI is 1 mas 
for M87 at distance $10$--$100$ mas from the central engine at frequency 15GHz 
(Yu. Kovalev, personal communication). Nevertheless, the core can be resolved 
with VLBI if the jet is less magnetised, e.g., for Michel magnetisation parameter 
$\sigma_{\rm M} =10$ (see Fig.~\ref{Doppler10}).

We also present the results for the distributions of the Doppler factor in the 
case of fixed angle between rotational axis of the jet and the line-of-sight
$\alpha=18^{\circ}$ (as for the jet in M87) and different magnetisation 
parameters $\sigma_{\rm M} = (10, 30)$. The value $\sigma_{\rm M} = 100$ 
was considered earlier. Larger values $\sigma_{\rm M} > 100$ look unreasonable 
because only the radiation from the very narrow ring at the jet boundary 
could be detected for large $\sigma_{\rm M}$, which does not match the 
observations of more wider structures across the M87 jet.

As was shown above, outer annular region of a jet has both low plasma 
density and low magnetic field pressure. Hence, radiation from this region
is to be double-depressed. The map of luminosity distribution will be presented 
in the future papers. Here for simplicity we consider everywhere the solutions
with the same radius $r_{\rm jet}=3.7$ pc corresponding to ambient pressure 
$P_{\rm ext}\approx 10^{-10} \mathrm{dyn/cm^2}$ for $\sigma_{\rm M} = 100$
and to $P_{\rm ext}\approx 10^{-9} \mathrm{dyn/cm^2}$ for 
$\sigma_{\rm M} = 10$ (see Fig. 5 in~\citealt{BCKN-17}). 

Finally, we discuss the polarisation properties of radio emission, which do not 
depend on the number density of radiating particles. Indeed, relativistic 
motion of emitting plasma also affects the direction of observed linear polarisation 
of synchrotron radiation which properties in the rest frame of plasma (where only 
ordered magnetic field ${\bf B}^{\prime}$ is present and electric field 
${\bf E}^{\prime}$ vanishes) are well known (see, e.g., chapter 5 in 
\citealt{Ginzburg_book_89}). Specifically, for highly relativistic radiating 
particles the electric field ${\bf e}^{\prime}$ of the wave is perpendicular 
to the local direction of the static magnetic field ${\bf B}^{\prime}$. 

The changes in polarisation properties of polarised electromagnetic wave under 
Lorentz transformations was first mentioned and applied in astrophysical settings 
by~\citet{cocke72}. After Lorentz boost the observed direction of the wave 
electric vector in observer's reference frame is, in general, not perpendicular 
to the direction of the magnetic field. Calculations of polarisation properties 
of synchrotron radiation emitted by relativistic extragalactic jets were done by
many authors (see, e.g.,~\citealt{BlandfordKonigl79, Parievetal, LPG, PerRom18}). 
However, two dimensional distribution of polarisation over cross section of the 
jets was not included into consideration in these works. 

Here we calculate and draw maps of unit vector ${\hat{\bf e}}$ along the wave electric 
field ${\bf e}$ in linearly polarised synchrotron radiation as seen by the observer. 
Each small patch of plasma contains isotropically distributed relativistic particles
with energy spectrum $dN = K \epsilon^{-p} d\epsilon$. We do not consider circular
polarisation at the moment. This approximation  corresponds to ultrarelativistic 
energies of emitting particles. Then, the degree of linear polarisation emitted by 
every small patch of the jets is $\Pi_0 = (p+1)/(p+7/3)$~\citep{Ginzburg_book_89}. 
Under vacuum approximation the observed radiation is obtained by integration of Stokes 
parameters of independent incoherent emitters over the line of sight. Because each 
emitter has varying direction of the magnetic field and varying relativistic velocities, 
the direction of polarisation ${\hat{\bf e}}$ for each emitter also varies. As a result, 
the degree of polarisation in the total integrated radiation flux along each line of 
sigh will be smaller than the upper limit $\Pi_0$. 

Here we leave the construction of integrated observable maps of synchrotron radiation 
(including consideration of propagation effects in thermal plasma, such as Faraday 
rotation, as well as self-absorption effects) for forthcoming series of works. Below 
we restrict ourselves only to the construction of polarisation maps that correspond 
to individual cross sections of a jet for given distance from the ''central engine''.

General expressions giving the polarisation unit vector ${\hat {\bf e}}$ in terms of 
the observed direction of magnetic field at the emitter, unit vector ${\hat {\bf B}}$,
direction of the wave vector of the wave to the observer, unit vector ${\hat {\bf n}}$,
and the velocity ${\bf v}$ of the emitter were derived in compact form in \citet{LPB03}, 
formula (C5) in Appendix C there. We reproduce these expression here for convenience 
and keep speed of light $c$ in line with the notations used in the present work:
\begin{equation}
{\hat {\bf e}} ={  {\bf n} \times {\bf q} \over 
\sqrt{ q^2 - ( {\bf n} \cdot {\bf q})^2} } \mbox{,}
\quad
{\bf q} = {\hat{\bf B}} +
 {\bf n}  \times (  {\bf v} \times  {\hat{\bf B}})/c  \mbox{.}
\label{eee}
\end{equation}
Directions of polarisation vector ${\hat {\bf e}}$ are plotted in
Figs~\ref{Doppler18}--\ref{Doppler30} on panels~(b). As we see, 
linear polarisation has a rather complicated structure, which 
should be taken into account when analysing observations. 
In the observer's frame the magnetic field is dominated by the toroidal component for all radii, therefore, vectors ${\hat{\bf e}}$ would be directed radially everywhere on the map if the bulk motion is non-relativistic. In reality the flow is non-relativistic only in the vicinity of the outer boundary of the jet, because at the boundary itself all components of velocity vanish. Accordingly,
we observe the ring of radially directed polarisation vectors near the outer edge of the jet on panels~(b) in all Figs~\ref{Doppler18}--\ref{Doppler30}. 
Inside the ring the swing of the polarisation direction due to the relativistic aberration becomes significant when $\gamma(r) \alpha \geq 1$. Because $\gamma$ decreases toward the jet boundary, the size of the ring with radially directed polarisation vectors ${\hat{\bf e}}$ is smaller for larger $\alpha$ and is also smaller for faster flows with larger $\gamma$ and larger $\sigma_{\rm M}$.

\section{Discussion and conclusion}

In this paper we investigated the internal structure of relativistic 
jets and their influence on the polarisation properties of the 
observed radiation. In particular, the profiles of magnetic field, 
velocity, and  number density across cylindrical jet submerged into 
non-magnetised gas at rest were determined. Another result is the map 
of the Doppler boosting factor at the cross-section of the jet together 
with the map of relativistic beaming effect under consideration.

It is shown how both the magnetic field and the number density gradually 
drop to external medium values close to the external boundary of the jet.  
We demonstrated that regardless the inclination angle $\alpha$ between 
the jet axis and the line-of-sight, only outer ring and central core can 
be observed. The size of central core is very small and can be measured 
only when the angle between the jet axis and the line-of-sight is small. 

It is necessary to notice that the central core must exist for MHD mechanism 
of acceleration since the electromagnetic energy flux $\sim \Omega_{\rm F} I$ 
at the rotation axis is zero, and Lorentz factor is conserved along the axis.
The outer part may be relatively large with dramatic change of Doppler factor 
across the ring, but the outer part is supposed to be dim.

The multiplicity parameter calculated using results of our model is
\begin{equation}
\lambda=\frac{n^{\rm (lab)}}{n_{\rm GJ}} \approx 5\times10^{12},\label{nngj}
\end{equation}
where $n_{\rm GJ}=\Omega_{\rm F} B_{\rm p}/(2 \pi {\rm c}  {\rm e}) $ is Goldreich-Julian
density, which is the number density of charged particles just enough to screen the
longitudinal electric field. This result is in agreement with observations because 
$\lambda \sigma_{\rm M} \approx (W_{\rm tot}/W_{\rm a})^{1/2}$, where $W_{\rm tot}$ is 
the total energy losses of the jet, and \mbox{$W_{\rm a}= m_{\rm e}^2 c^5/e^2
\approx 10^{17} {\rm erg/s}$} (see~\citealt{MHD} for more detail). It gives 
$\lambda \sigma_{\rm M} \sim 10^{14}$, which agrees with~(\ref{nngj}) for 
$\sigma_{\rm M}$ from $10$ to $100$.

In addition, it is shown that the size of the central core actually does not depend 
on Michel magnetisation parameter $\sigma_{\rm M}$ but is a function of the inclination 
angle $\alpha$. In contrast, the size of the outer ring depends on both $\sigma_{\rm M}$ 
and the angle $\alpha$: the lower each value, the wider the ring. The most reasonable 
Michel parameter to explain observations of M87 is $\sigma_{\rm M} \approx 10$. This 
value of $\sigma_{\rm M}$ is able to reproduce simultaneously the size of the outer
bright core, bulk Lorentz factor $\gamma \approx 6$, the observation of superluminal
motion, and numerical simulations (see, e.g.,~\citealt{Porth_etal11}). 
 We have also calculated observed directions of linear polarisation of 
synchrotron radiation emitted by cross sectional layers of the jet. The effect of 
relativistic aberration on polarisation was also taken into account. This 
effect and obtained distributions of polarisation direction are fundamental basics of 
understanding and interpreting present and future VLBI polarisation measurements
of relativistic magnetised jets. Full integrated radio images of the jet together with its 
rotation measure and polarisation will be presented in the following papers. 

\section{Acknowledgments}

We thank Y.N.~Istomin, S.V.~Chernov, and E.E.~Nokhrina for useful discussions, and especially 
14$^{\rm th}$ School of Modern Astrophysics SOMA-2018 (\url{http://astrosoma.ru}) for
provision of fruitful ideas and discussions. This work was supported by 
Russian Foundation for Basic Research (Grant no. 17-02-00788).

\bibliography{mybib}
\bibliographystyle{mnras}

\label{lastpage}

\end{document}